\documentclass[12pt]{article}
\usepackage[dvips]{graphicx}

\usepackage{epsfig,amsmath,amssymb,verbatim,mathrsfs}
\def\beq{\begin{eqnarray}}
\def\eeq{\end{eqnarray}}
\def\bea{\begin{eqnarray}}
\def\eea{\end{eqnarray}}

\def\tev{\, {\rm TeV}}
\def\gev{\, {\rm GeV}}

\newcommand{\gsim}{\lower.7ex\hbox{$\;\stackrel{\textstyle>}{\sim}\;$}}
\newcommand{\lsim}{\lower.7ex\hbox{$\;\stackrel{\textstyle<}{\sim}\;$}}

\def\stilde{\widetilde}

\newcommand{\newc}{\newcommand}
\newc{\Nc}{N_{c}}
\newc{\CG}{C_G}
\newc{\gp}{g'}
\newc{\stopi}{\stilde t_i}
\newc{\sboti}{\stilde b_i}
\newc{\staui}{\stilde \tau_i}
\newc{\stopj}{\stilde t_j}
\newc{\sbotj}{\stilde b_j}
\newc{\stauj}{\stilde \tau_j}
\newc{\stopI}{\stilde t_1}
\newc{\stopII}{\stilde t_2}
\newc{\sbotI}{\stilde b_1}
\newc{\sbotII}{\stilde b_2}
\newc{\stauI}{\stilde \tau_1}
\newc{\stauII}{\stilde \tau_2}
\newc{\sstop}{s_{t}}
\newc{\cstop}{c_{t}}
\newc{\ssbot}{s_{b}}
\newc{\csbot}{c_{b}}
\newc{\sstau}{s_{\tau}}
\newc{\cstau}{c_{\tau}}
\newc{\Sstop}{s_{2t}}
\newc{\Cstop}{c_{2t}}
\newc{\Ssbot}{s_{2b}}
\newc{\Csbot}{c_{2b}}
\newc{\Sstau}{s_{2\tau}}
\newc{\Cstau}{c_{2\tau}}
\newc{\salpha}{s_\alpha}
\newc{\calpha}{c_\alpha}
\newc{\Calpha}{c_{2\alpha}}
\newc{\Salpha}{s_{2\alpha}}
\newc{\sbetapm}{s_{\beta_\pm}}
\newc{\cbetapm}{c_{\beta_\pm}}
\newc{\Sbetapm}{s_{2 \beta_\pm}}
\newc{\Cbetapm}{c_{2 \beta_\pm}}
\newc{\sbetaO}{s_{\beta_0}}
\newc{\cbetaO}{c_{\beta_0}}
\newc{\SbetaO}{s_{2 \beta_0}}
\newc{\CbetaO}{c_{2 \beta_0}}
\newc{\vu}{v_u}
\newc{\vd}{v_d}
\newc{\seL}{\stilde e_L}
\newc{\smuL}{\stilde \mu_L}
\newc{\seR}{\stilde e_R}
\newc{\smuR}{\stilde \mu_R}
\newc{\suL}{\stilde u_L}
\newc{\sdL}{\stilde d_L}
\newc{\suR}{\stilde u_R}
\newc{\sdR}{\stilde d_R}
\newc{\scL}{\stilde c_L}
\newc{\ssL}{\stilde s_L}
\newc{\scR}{\stilde c_R}
\newc{\ssR}{\stilde s_R}
\newc{\snue}{\stilde \nu_e}
\newc{\snumu}{\stilde \nu_\mu}
\newc{\snutau}{\stilde \nu_\tau}
\newc{\Gpm}{G^\pm}
\newc{\Hpm}{H^\pm}
\newc{\FFbS}{\overline{FF}S}
\newc{\FFbV}{\overline{FF}V}
\newc{\FSS}{F_{SS}}
\newc{\FSSS}{F_{SSS}}
\newc{\FFFS}{F_{FFS}}
\newc{\FFFbS}{F_{\overline{FF}S}}
\newc{\FSSV}{F_{SSV}}
\newc{\FVS}{F_{VS}}
\newc{\FVVS}{F_{VVS}}
\newc{\FFFV}{F_{FFV}}
\newc{\FFFbV}{F_{\overline{FF}V}}
\newc{\Fgauge}{F_{\rm gauge}}
\newc{\DRbarprime}{$\overline{\rm DR}'$ }
\newc{\DRbar}{$\overline{\rm DR}$ }
\newc{\MSbar}{$\overline{\rm MS}$ }
\newc{\Yu}{{\bf Y}_u}
\newc{\Yd}{{\bf Y}_d}
\newc{\Ye}{{\bf Y}_e}
\newc{\Au}{{\bf a}_u}
\newc{\Ad}{{\bf a}_d}
\newc{\Ae}{{\bf a}_e}
\newc{\bm}{{\bf m}}
\newc{\zhol}{Z^{\rm hol}}

\newc{\rwino}{r_{\tilde W}}
\newc{\rmu}{r_{\tilde H}}
\newc{\ra}{r_A}
\newc{\ccdot}{\!\cdot\!}

\newcommand{\nnmb}{\nonumber}

\newcommand{\lrf}[2]{\left(\frac{#1}{#2}\right)}

\newcommand\T{\rule{0pt}{2.6ex}}
\newcommand\B{\rule[-1.2ex]{0pt}{0pt}}

\begin{document}

\setlength{\baselineskip}{0.2in}

%\begin{comment}

%@twocolumnfalse\endcsname
%%
%%
\begin{titlepage}
\noindent
%\begin{flushright}
%\end{flushright}
%%
%%
%\vspace{1cm}

\begin{center}
  \begin{Large}
    \begin{bf}
Baryon Destruction by Asymmetric Dark Matter
     \end{bf}
  \end{Large}
\end{center}
\vspace{0.2cm}

\begin{center}

%\begin{large} - dm: I tried to get all the names on one line
Hooman Davoudiasl$^{(a)}$, David E. Morrissey$^{(b)}$,
Kris Sigurdson$^{(c)}$, Sean Tulin$^{(b)}$\\
%\end{large}
\vspace{1cm}
  \begin{it}
$(a)$ Department of Physics, Brookhaven National Laboratory,\\
Upton, NY 11973, USA\\
\vspace{0.2cm}
$(b)$ Theory Group, TRIUMF, \\
4004 Wesbrook Mall, Vancouver, BC V6T 2A3, Canada\\
\vspace{0.2cm}
$(c)$ Department of Physics and Astronomy, University of British Columbia,\\
Vancouver, BC V6T 1Z1, Canada\\
\vspace{0.5cm}
email: hooman@bnl.gov, dmorri@triumf.ca, krs@physics.ubc.ca,
tulin@triumf.ca
\end{it}

\end{center}

\center{\today}

\begin{abstract}

We investigate new and unusual signals that 
arise in theories where dark matter is asymmetric and carries a 
net antibaryon number, as may occur when the dark matter abundance 
is linked to the baryon abundance.
Antibaryonic dark matter can cause {\it induced nucleon decay}
by annihilating visible baryons
through inelastic scattering.
These processes lead to an effective nucleon lifetime of
$10^{29}-10^{32}$ years in terrestrial nucleon decay experiments,
if baryon number transfer between visible and dark sectors arises through
new physics at the weak scale.  The possibility of induced nucleon decay
motivates a novel approach for direct detection of cosmic dark matter
in nucleon decay experiments.
Monojet searches (and related signatures) at hadron colliders 
also provide a complementary probe of weak-scale dark-matter--induced 
baryon number violation.
%
%dm mod (I think we should keep the qualifier here)
Finally, we discuss the effects of baryon-destroying dark matter
on stellar systems and show that it can be consistent
with existing observations.
%
% Finally,
% %Additionally, 
% we discuss the implications 
% % KS mod
% for this type
% % KS mod
% of dark matter capture and baryon
% destruction in stellar environments and show
% % that it can be consistent
% it is consistent
% with existing observations.

 \end{abstract}

\vspace{1cm}

\end{titlepage}

\setcounter{page}{2} %so that the .pdf file numbering matches the labels.

%%%%%%%%%%%%%%%%%%%%%%%%%%%%%%%%%%%%%%%%%%%%%%%%%%%%%%%%%%%%%%%%%%%%%%

\section{Introduction\label{sec:intro}}

Cosmological observations indicate that about 4.6\% of the energy
density of the Universe consists of baryonic matter,
while 23\% is dark matter~(DM)~\cite{Komatsu:2010fb}.
Neither of these results can be explained with our current
understanding of elementary particles, the standard model~(SM).
Cosmology therefore requires
new fundamental physics, and it is important to find ways to detect 
such new physics experimentally.

In the majority of new-physics scenarios, the generation of baryons and DM
occurs through unrelated mechanisms, offering no explanation for the
similar magnitudes of their cosmological densities.  The most thoroughly
studied scenarios involve baryon production from CP-violating
non-equilibrium processes during the electroweak phase
transition, from decays of right-handed neutrinos, or from the coherent
evolution of scalar fields~\cite{Riotto:1999yt}, while the DM relic density
is determined by thermal freeze out when a non-relativistic stable species
falls out of equilibrium~\cite{Jungman:1995df}.
In this context, there is no reason to expect similar
cosmological densities of baryons and DM.

This apparent coincidence may instead be a clue that both
types of matter have a common origin.  Several models of asymmetric
dark matter~(ADM) have been proposed along these lines where the DM
density carries a net (approximately) conserved global charge shared
by the SM~\cite{Nussinov:1985xr,Hooper:2004dc,Kitano:2004sv,Agashe:2004bm,
Farrar:2005zd,Shelton:2010ta,
Hut:1979xw,Dodelson:1989cq,Kuzmin:1996he,Gu:2007cw,An:2009vq,
Davoudiasl:2010am},
such as baryon number $B$.
These models generally fall into two classes depending on how
the charge asymmetry is created:
\begin{enumerate}
\item An initial charge asymmetry, generated in either the visible
or DM sector, is partitioned between the two sectors by chemical
equilibration through a transfer operator~\cite{Nussinov:1985xr,
Shelton:2010ta}.
These charges are separately ``frozen in'' once the transfer operator
goes out of equilibrium.
\item Non-equilibrium dynamics
generate equal and opposite charge asymmetries in the visible
and DM sectors, without any net overall charge
asymmetry~\cite{Hut:1979xw,Dodelson:1989cq,Kuzmin:1996he,
Gu:2007cw,An:2009vq,Davoudiasl:2010am}.
In order to avoid washout, transfer operators must always be out
of equilibrium once the asymmetries are created.  We term this process
{\it hylogenesis} (``hyle'' = matter)~\cite{Davoudiasl:2010am}.
\end{enumerate}
Subsequently, in both cases, DM particles and antiparticles
are assumed to annihilate efficiently, leaving only a remnant
asymmetric component determined by the charge density.

  In the present work, we investigate novel experimental signatures from
hylogenesis scenarios where DM carries $B$~\cite{Davoudiasl:2010am}.
Here, the lowest-dimensional, gauge-invariant transfer operator is given by
\beq \label{INDop}
\mathscr{L}_{\textrm{eff}} \sim
\frac{1}{\Lambda^3} \, u_R^i d_R^j d_R^k \Psi_R \Phi + \textrm{h.c.}
\label{leff}
\eeq
where $i,j,k$ label generation.  In this case, DM has two components,
a fermion/scalar pair $(\Psi, \Phi)$
with total baryon number $B_{\Psi} + B_{\Phi} = -1$.  Stability of both
$\Psi$ and $\Phi$ requires $|m_\Psi - m_\Phi| < (m_p+m_e)$.
%*%
This operator is inactive cosmologically and does not wash out the
baryon-DM asymmetry for $\Lambda \gtrsim 100\,\gev$ provided the asymmetry
is created at relatively low temperatures, below about a GeV.
%*%
Our results may also be applicable to other ADM scenarios
in which the same transfer operator appears with one or both of
$(\Psi,\Phi)$ making up the DM.

  In hylogenesis, the Universe is net $B$-symmetric,
and therefore the baryon asymmetry carried by DM
%$(\Psi,\Phi)$
is equal and opposite to that in visible baryons.  Specifically,
in the hylogensis scenario of Ref.~\cite{Davoudiasl:2010am},
this has two important consequences:
\vspace{-0.3cm}
\begin{itemize}
\item The number densities of $(\Psi,\Phi)$ satisfy \mbox{$n_\Psi = n_\Phi = n_B$}.
Therefore, cosmological observations imply
$(m_\Psi + m_\Phi)/m_p = \Omega_{DM}/\Omega_b \approx 5$.  Together with the
DM stability requirement, we have $m_{\Psi,\Phi} \approx 1.7 - 2.9$ GeV.
\item There exist many scenarios in which ADM is coupled to the SM via the
``neutron portal'' operator $u_R d_R d_R$~\cite{Nussinov:1985xr},
generally falling into the first class of ``chemical equilibration''
scenarios.  In this case, DM is very often baryonic ($B_{DM} >0$),
while for hylogenesis scenarios the DM must be antibaryonic 
($B_{DM} < 0$).\footnote{
In supersymmetric models, light superpartners can play an
important role in chemical equilibration~\cite{Chung:2008gv}, potentially
affecting the sign of $B_{DM}$ in equilibration ADM scenarios.
Also, for other applications of the neutron portal operator
to baryogenesis, see Ref.~\cite{Dimopoulos:1987rk}.
}
\end{itemize}
\vspace{-0.3cm}
In Ref.~\cite{Davoudiasl:2010am}, we presented a specific realization for
hylogenesis where the operator in Eq.~\eqref{INDop} arises by
integrating out heavy Dirac fermion mediatiors $X_{1,2}$.
Out-of-equilibrium decays of $X_1$ during reheating generate
equal-and-opposite dark and visible baryon asymmetries.
Furthermore, in this scenario $(\Psi,\Phi)$ are charged
under an additional hidden $U(1)'$ gauge symmetry,
that couples to the SM via kinetic mixing with hypercharge,
to facilitate annihilation of the symmetric DM densities.
More details about this realization are given in Appendix~\ref{sec:appa}.

Hylogenesis models with the operator of Eq.~\eqref{INDop}
have an interesting and unique signature:
antibaryonic DM particles can annihilate visible baryonic matter, termed
{\it induced nucleon decay} (IND).
IND is a novel and unusual DM signal.  These events are
$\Psi N \to \Phi^\dagger M$ and $\Phi N \to \bar{\Psi} M$, where
$N=(n,p)$ is a nucleon and $M$ is a meson.  Since the DM states
are invisible, this mimics nucleon decay, with an effective lifetime
dependent on the local DM density.
This signal offers the new and exciting possibility of
searching for DM in nucleon decay searches in
deep underground detectors such as
SuperKamiokande~\cite{Kobayashi:2005pe}.

In Section~\ref{sec:ind}, we compute the rates and kinematics of IND,
and discuss the implications of IND for existing nucleon decay searches.
Our main conclusions are: (i) due to different kinematics of IND, compared to
standard nucleon decay, existing bounds do not apply over most of the region
of parameter space of our DM model, and (ii) the effective nucleon lifetime
can be around $10^{29} - 10^{32}$ years, if the new physics scale $\Lambda$
in Eq.~\eqref{INDop} is the weak scale.

Hadron colliders can probe (anti)baryonic DM scenarios through direct
production of the weak-scale mediators transfering $B$ between visible and dark
sectors, discussed in Section~\ref{sec:coll}.
The operator of Eq.~\eqref{INDop} can give rise to observable
signatures in the form of monojets (or jets and missing energy).
Within our hylogenesis model, if IND were detected in nucleon decay searches,
monojet signals are inescapable and should be observed at
the Large Hadron Collider (LHC).

IND processes can potentially be relevant in stars.  Since the IND rate scales
with the DM density, capture and accretion of DM in stars is
important, possibly leading to modification of stellar evolution due to
baryon destruction and energy injection.  In Section~\ref{sec:star}, we consider
these effects in neutron stars, white dwarfs, and main-sequence stars.
Our conclusions are summarized in Section~\ref{sec:conc}.

\section{Dark Matter Detection in Nucleon Decay Searches
\label{sec:ind}}

%---------------------------------------------------------
\begin{figure}[ttt]
\begin{center}
\includegraphics[scale=0.95]{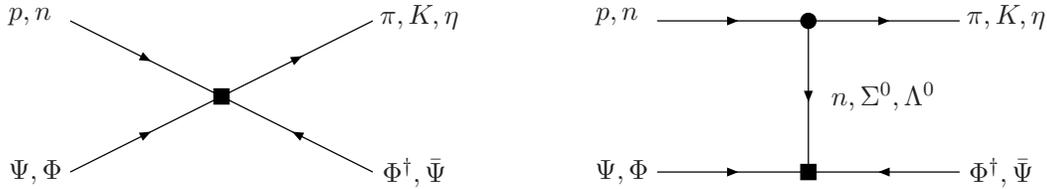}
\end{center}
\vspace{-.4cm}
\caption{\it \small  Feynman diagrams for induced nucleon decay.
Box denotes IND vertex from Eq.~\eqref{eq:Leff}. Circle denotes strong interaction
vertex given by $\mathscr{L}_0$ in Ref.~\cite{Claudson:1981gh}. }
\label{indfeyn}
\end{figure}
%---------------------------------------------------------

Dark matter particles $(\Psi, \Phi)$ can annihilate nucleons $N$, producing an energetic meson $M$ through inelastic scattering
\beq \label{INDprocess}
\Phi N \longrightarrow \bar{\Psi} M \; , \qquad \Psi N \longrightarrow \Phi^\dagger M \;,
\eeq
shown in Figure~\ref{indfeyn}.  We restrict our attention to single meson final states, although
multi-meson events are allowed and may have comparable rates.  In general, both down-scattering
and up-scattering can occur, defined as whether the heavier or lighter DM particle, respectively,
is in the initial state; however, the latter is kinematically forbidden if $|m_\Psi - m_\Phi| > m_N - m_M$.
Since neither the initial DM nor final anti-DM particles are observed directly, these processes mimic standard
nucleon decay events $N \to M \nu$ \cite{Nath:2006ut},
with an undetected final state neutrino $\nu$ (or antineutrino $\bar{\nu}$).

The observable meson energy for each case in Eq.~\eqref{INDprocess} is,
respectively,
\beq
E_M = \frac{(m_N + m_\Phi)^2 + m_M^2 - m_\Psi^2}{2(m_N+m_\Phi)} \, , \quad
E_M = \frac{(m_N + m_\Psi)^2 + m_M^2 - m_\Phi^2}{2(m_N+m_\Psi)}  \;  ,
\eeq
neglecting the initial kinetic energy of the DM particles ($v_{DM} \sim 10^{-3}$).

Although hylogenesis (as a baryogenesis mechanism) works for any quark flavor,
the most interesting signatures arise if the IND operators involve $u,d,s$ quarks only.
Restricting our attention to the lightest mesons ($\pi$, $K$, $\eta$), the final state meson
can have much more kinetic energy in IND than for standard nucleon decay (SND), summarized in Table.~\ref{tab:INDmom}.
For fixed masses $m_{\Psi}$ and $m_{\Phi}$, the meson is either monochromatic or bichromatic (up to Fermi motion),
depending on whether up-scattering is forbidden or allowed.  The range of momenta $p_M$ corresponds to the allowed range
$m_{\Psi,\Phi} \approx 2-3$ GeV, provided that $(\Psi,\Phi)$ are stable and account for the observed $\Omega_{DM}$.

%%%%%%%%%%%%%%%%%%%%% Table    %%%%%%%%%%%%%%%%%
\begin{table}[ttt]
\begin{center}
\begin{tabular}{|c|c|c|c|l|}
\hline
Decay mode \T \B & $p_M^{SND}$ & $p_M^{IND}$ [up]& $p_M^{IND}$ [down] & $\tau_N^{SND}$ bound ($\times 10^{32}$ yr)\\
\hline
$N \to \pi$ \T\B& 460 & $< 800$ & $800 - 1400$ & $\tau_p^{SND} > 0.16 $~\cite{Wall:2000pq} , \,  $\tau_n^{SND} > 1.12$~\cite{McGrew:1999nd}\\
$N \to K$ \B & 340 & $< 680$ & $680 - 1360$ & $\tau_p^{SND} > 23$~\cite{Kobayashi:2005pe} , \;\;\;\quad  $\tau_n^{SND} > 1.3$~\cite{Kobayashi:2005pe}\\
$N \to \eta$ \B & 310 & $< 650$ & $650 - 1340$ & $\tau_n^{SND} > 1.58 $~\cite{McGrew:1999nd}\\
\hline
\end{tabular}
\end{center}
\caption{\it \small Comparison of meson $M= (\pi, K, \eta)$ momentum $p_M$ (MeV) for standard nucleon decay (SND) and induced nucleon decay (IND) from DM, for
up- and down-scattering.}
\label{tab:INDmom}
\vspace{-0.5cm}
\end{table}
%%%%%%%%%%%%%%%%%%%%%%%%%%%%%%%%%%%%%%%%%%%

\subsection{Nucleon decay searches}

Existing searches have been optimized for meson momenta $p_M^{SND}
\sim 300 - 450$ MeV, while for IND mesons are typically much more
energetic, with momenta $p_M^{IND} \sim 1$ GeV.  Here, we briefly
summarize existing SND search strategies and how they might be
adapted for IND searches.

\noindent{$\underline{p \to K^+ \nu, \; n \to K^0 \nu}${\bf :} }
The Super-Kamiokande experiment,
a water \v{C}erenkov detector, provides the strongest limits on these
channels.  For $K^+$, they have three searches: ({\it i}) $K^+ \to
\pi^+ \pi^0$, giving three \v{C}erenkov rings, ({\it ii}) $K^+ \to
\mu^+$ with a prompt $\gamma$ (from $^{16}\textrm{O} \, \to \,
^{15}\textrm{N}^* \, \to \, ^{15}\textrm{N} + \gamma$), and ({\it
iii}) a mono-energetic $\mu^+$ from $K^+ \to \mu^+$, with no prompt
$\gamma$.  All three searches assume, as is the case for $p_{K^+}
\approx 340$ in SND, that the $K^+$ is emitted below \v{C}erenkov
threshold ($\beta < 0.75$) and comes to rest before decaying.  For
IND, we estimate that an $\mathcal{O}(1)$ fraction of $K^+$'s do
come to rest before decaying.  However, except for up-scattering
events close to kinematic threshold, the $K^+$ from IND has $\beta >
0.75$, adding an extra ring to the event topology.  Furthermore,
this additional radiation may make finding the prompt $\gamma$ in
search method ({\it ii}) more difficult.  For $K^0$, they have two
searches: ({\it iv}) $K^0_S \to \pi^0 \pi^0 \to 4 \gamma$, giving
four $e$-like rings, and ({\it v}) $K^0_S \to \pi^+ \pi^-$, giving
two $\mu$-like rings.  Both searches assume $200 < p_{K^0} < 500$
MeV, thereby excluding IND events (again, except for up-scattering
near kinematic threshold).  One difficulty in search ({\it iv}) is
identifying all four $e$-like rings.  For IND, this may be more
difficult as the rings would be more overlapping due to relativistic
beaming.  On the other hand, search ({\it v}) is promising for IND.
In SND this mode suffers from a small efficiency that the $\pi^\pm$
are both above \v{C}erenkov threshold.  With greater energetics in IND,
the efficiency may be much larger.

\noindent $\underline{p \to \pi^+ \nu}${\bf :}
The best limit is provided by the Soudan 2
experiment, an iron tracking calorimeter.    Nucleon decay event
candidates were required to have a single $\pi^+$ track, with
ionization consistent with mass $m_\pi$ or $m_\mu$, initial momentum
$140 < p_{\pi^+} < 420$ MeV, and visible endpoint decays $(\pi^+ \to
\mu^+ \to e^+$).  Their simulations showed that a $\pi^+$
originating from within an iron nucleus loses on average half its
initial momentum.  At higher $p_{\pi^+}$, IND events may be more
visible due to reduced background from atmospheric neutrinos.
However, it is unknown to us what is the average momentum deposition
in iron of the $\pi^+$ at much higher energy, and whether this can
lead to fragmentation of the parent nucleus into exotic nuclear
states.

\noindent$\underline{n \to \pi^0 \nu, \; n \to \eta \nu}${\bf :}
The best limits on these
modes come from the IMB-3 experiment, a water \v{C}erenkov
detector.\footnote{We note that the IMB-3 experiment found an excess
in events with total energy $900-1100$ MeV (20 events vs.~6.1
expected background)~\cite{McGrew:1994nd}. Nearly all these events
had between $2-4$ \v{C}erenkov rings, large missing momenta ($400 -
1100$ MeV), and large invariant masses ($600 - 1100$ MeV).  In IND,
a large missing momentum would be expected, while
a large invariant mass could arise through heavy meson or
multi-meson final states.}  The $\pi^0 \to \gamma \gamma$ channel
may be more difficult at higher energies: due to decreased
separation angle of the two photons (from relativistic beaming),
they can appear as a single electron-like track.  The $n \to \eta
\nu, \; \eta \to \gamma \gamma$ channel will have greater photon
separation and may be more promising.  We find that the IND rates
into $\pi^0$ and $\eta$ final states are comparable and are
sensitive to the same underlying IND operator (shown below).

\subsection{Effective nucleon lifetime from IND}

An effective IND lifetime can be defined as the inverse scattering
rate per target nucleon, $\tau_{N}^{-1} \equiv n_{DM} (\sigma
v)_{IND}$, with local DM number density $n_{DM}\equiv
\rho_{DM}/(m_\Psi + m_\Phi)$ and IND scattering cross section
$(\sigma v)_{IND}$. Numerically, we have \beq \tau_{N}^{-1} \approx
(10^{32} \; \textrm{yrs})^{-1} \times \left(\frac{\rho_{{DM}}}{0.3
\, \textrm{GeV}/\textrm{cm}^3} \right) \left(\frac{(\sigma
v)_{{IND}} }{10^{-39} \, \textrm{cm}^3/\textrm{s}} \right) \; . \eeq

Next, we compute $(\sigma v)_{IND}$ using chiral perturbation
theory.  We perform an expansion in powers of $p_M/(4\pi f)$, where
$f \approx 139$ MeV is the pion decay constant, and truncate at
leading order.  Since for IND we expect $p_M \sim 4\pi f \sim 1\;
\textrm{GeV}$, our calculations should be regarded as
order-of-magnitude estimates at best.  Our analysis closely follows
SND rate computations in Ref.~\cite{Claudson:1981gh}.

There are four effective interactions that are relevant for IND processes
with single meson final states.  These are given by $\mathscr{L}_{\textrm{int}} = \sum_i c_i O_i$, with operators (given in two-component spinor notation)
\begin{align}
O_1 &= \epsilon_{\alpha\beta\gamma} \Phi (u^\alpha_R d_R^\beta)(d_R^\gamma \Psi_R) \\
O_2 &= \frac{1}{\sqrt{6}} \, \epsilon_{\alpha\beta\gamma} \Phi [  (d^\alpha_R s_R^\beta)(u_R^\gamma \Psi_R) + (s^\alpha_R u_R^\beta)(d_R^\gamma \Psi_R)
- 2 (u^\alpha_R d_R^\beta)(s_R^\gamma \Psi_R)] \\
O_3 &= \frac{1}{\sqrt{2}} \, \epsilon_{\alpha\beta\gamma} \Phi [ (d^\alpha_R s_R^\beta)(u_R^\gamma \Psi_R)  - (s^\alpha_R u_R^\beta)(d_R^\gamma \Psi_R) ]
\end{align}
where $\alpha,\beta,\gamma$ are color indeces, and the coefficients $c_i$ have mass dimension $-3$.\footnote{A fourth operator $\epsilon_{\alpha\beta\gamma} \Phi (s_R u_R)(s_R\Psi_R) $ is relevant only for multi-kaon final states.  A fifth operator $\epsilon_{\alpha\beta\gamma} \Phi [ (d^\alpha_R s_R^\beta)(u_R^\gamma \Psi_R) + (s^\alpha_R u_R^\beta)(d_R^\gamma \Psi_R) + (u^\alpha_R d_R^\beta)(s_R^\gamma \Psi_R)]$ vanishes by a Fierz identity.}
The linear combinations have been chosen such that $O_{1,2,3}$ have strong isospin $I=(\frac{1}{2}, \, 0, \, 1)$, respectively.  Here, it is useful to write $\mathscr{L}_{\textrm{int}} = \textrm{Tr}(c \,O)$ where
\beq
c \equiv \left( \begin{array}{ccc} \frac{c_2}{\sqrt{6}} + \frac{c_3}{\sqrt{2}} & 0 & 0 \\ 0 & \frac{c_2}{\sqrt{6}} - \frac{c_3}{\sqrt{2}} & 0 \\ 0 & c_1 & - \sqrt{\frac{2}{3}} \, c_2 \end{array} \right) \, , \quad
O_{ij} \equiv \frac{1}{2}\, \epsilon_{\alpha\beta\gamma} \,{\epsilon_{jk\ell}} \, (q_{Rk}^\alpha q_{R\ell}^\beta) (q^\gamma_{iR} \Psi_R)  \Phi \; ,
\eeq
with $q_R^\alpha \equiv (u,d,s)^\alpha_R$.  Under $SU(3)_L \times SU(3)_R$ chiral symmetry transformations, the right-handed quark fields transform in the $(\boldsymbol{1},\boldsymbol{3})$ representation, $q_R \to R \, q_R$ (where $R \in SU(3)_R$), while the IND operator $O_{ij}$ transforms in the $(\boldsymbol{1},\boldsymbol{8})$ representation, $O \to R O R^\dagger$.  If we treat $c$ as a spurion in the $(\boldsymbol{1},\boldsymbol{8})$ representation, $\mathscr{L}_{\textrm{int}}$ is invariant under chiral transformations.  (The DM fields $(\Psi,\Phi)$ are chiral singlets.)

The IND interactions of baryon and (pseudo-Goldstone) meson fields are determined by the chiral transformation properties of the spurion $c$.  Following the conventions of Ref.~\cite{Claudson:1981gh}, the only invariant operator is
\beq \label{eq:Leff}
\mathscr{L}_{IND} = \beta \, \textrm{Tr}[ c \, \xi^\dagger (B_R \Psi_R) \Phi \xi] \; ,
\eeq
where $\xi \equiv \exp(i M/f)$.  The meson and baryon fields are
\beq
M = \left( \begin{array}{ccc} \frac{\eta}{\sqrt{6}} + \frac{\pi^0}{\sqrt{2}}& \pi^+ & K^+ \\
\pi^- & \frac{\eta}{\sqrt{6}} - \frac{\pi^0}{\sqrt{2}}  & K^0 \\
K^- & \bar{K}^0 & -\sqrt{\frac{2}{3}} \, \eta \end{array} \right) \, , \quad
B = \left( \begin{array}{ccc} \frac{\Lambda^0}{\sqrt{6}} + \frac{\Sigma^0}{\sqrt{2}}   & \Sigma^+ & p \\
\Sigma^- & \frac{\Lambda^0}{\sqrt{6}} - \frac{\Sigma^0}{\sqrt{2}} & n \\
\Xi^- & \Xi^0 & -\sqrt{\frac{2}{3}}\, \Lambda^0 \end{array} \right) \; .
\eeq
Eq.~\eqref{eq:Leff} is invariant since the quantity $(\xi^\dagger B_R \xi)$ is in the $(\boldsymbol{1},\boldsymbol{8})$ representation~\cite{Claudson:1981gh}.  The unknown overall coefficient $\beta = 0.014(1) \; \textrm{GeV}^3$ has been computed using lattice methods~\cite{Aoki:1999tw}.

The Feynman diagrams for IND are given in Fig.~\ref{indfeyn}.  It is straight-forward to derive the Feynman rules for the interactions of baryons, mesons, and DM by expanding the matrix expressions in Eq.~\eqref{eq:Leff} and working to linear order in $1/f$.    From these, we compute the matrix elements and cross sections for IND processes.  We assume that only one coefficient $c_i$ is non-zero at a time.  $N \to \pi,\eta$ modes depend only on $c_1$, while $N \to K$ modes are governed by $c_{2,3}$.

%---------------------------------------------------------
\begin{figure}[ttt]
\begin{center}
\includegraphics[scale=0.6]{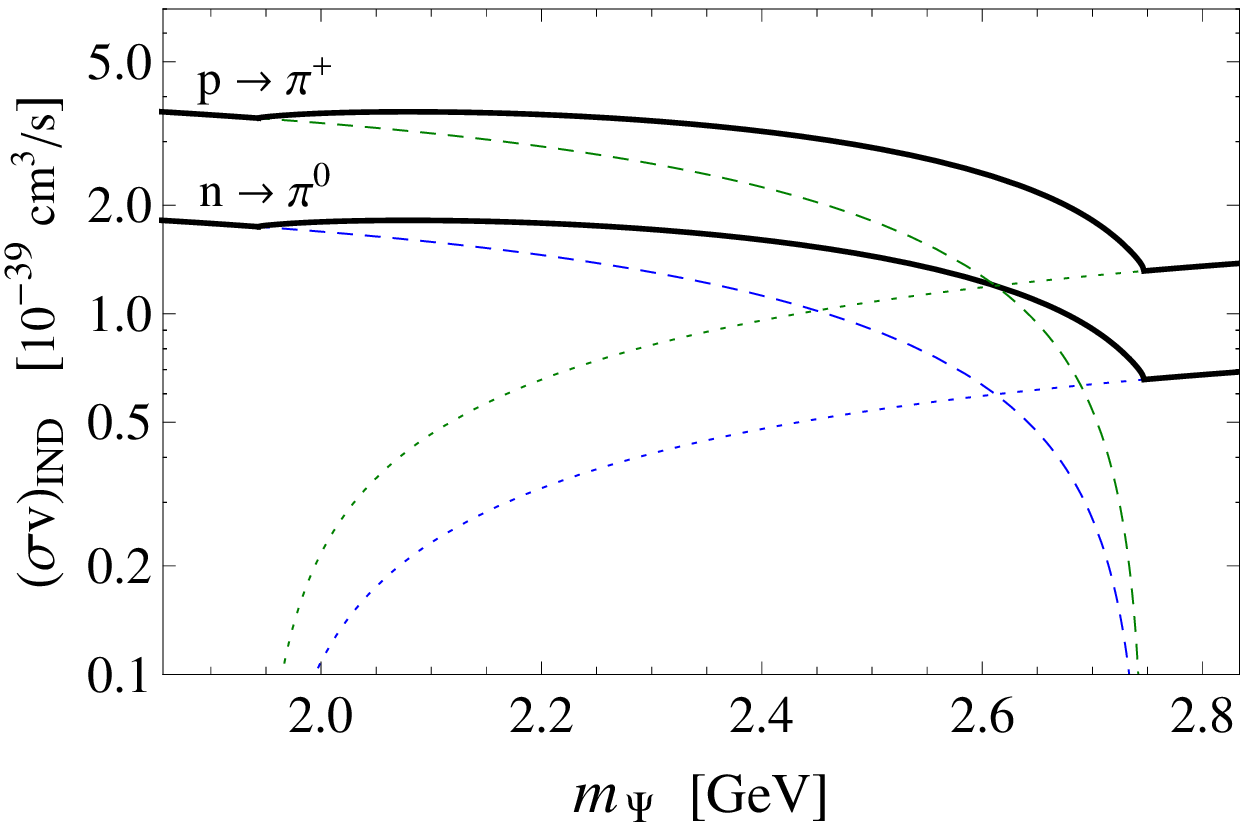} \quad \includegraphics[scale=0.6]{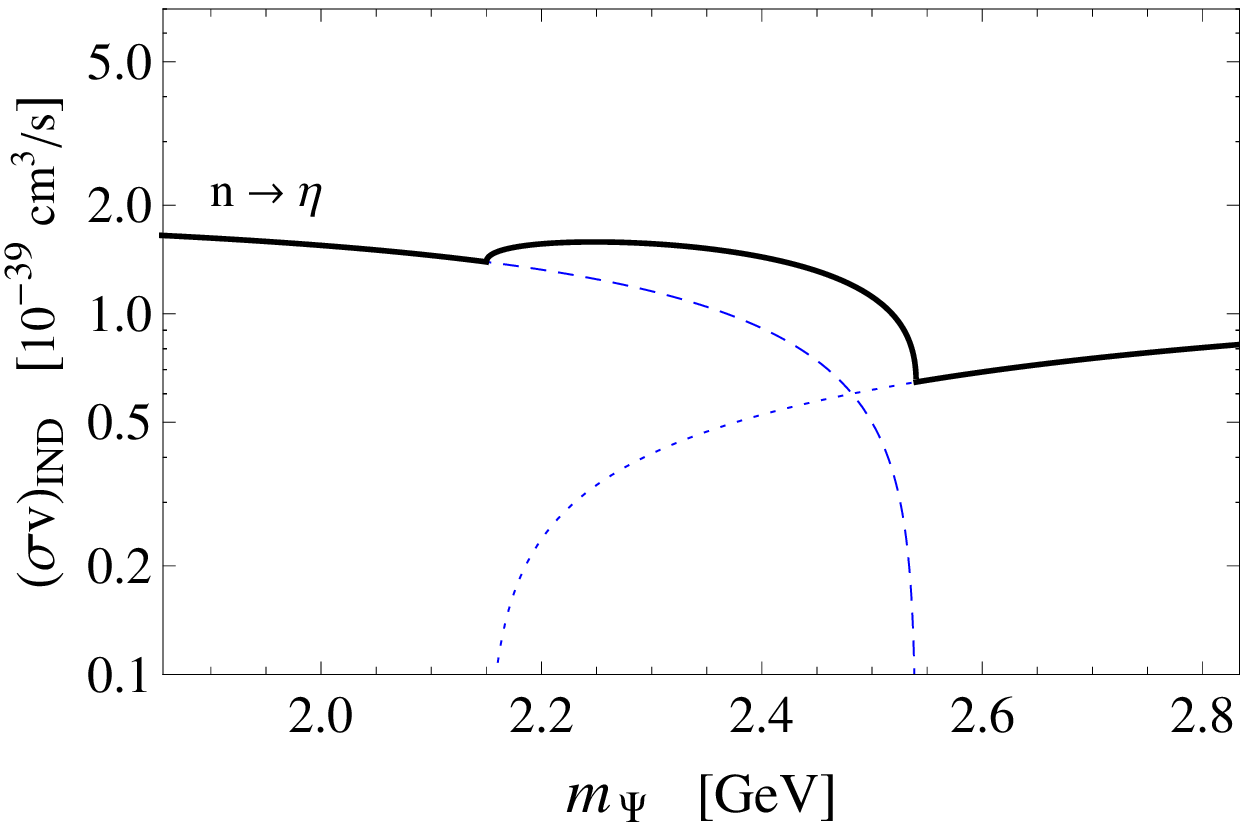}
\end{center}
\vspace{-.4cm}
\caption{\it \small  Induced nucleon decay cross sections $(\sigma v)_{IND}$ for $p,n \to \pi^+,\pi^0$ (left) and $n \to \eta$ (right) as a function of fermion DM mass $m_{\Psi}$ for $|c_1| =\textrm{TeV}^{-3}$.  Dotted (dashed) lines denote $N \Phi \to \bar{\Psi} M$ ($N \Psi \to \Phi^\dagger M$).  Solid lines denote total rates $N \Phi \to \bar{\Psi} M$ + $N \Psi \to \Phi^\dagger M$.  $(\sigma v)_{IND} = 10^{-39}$ cm$^3$/s corresponds to lifetime $\tau_N^{IND} = 10^{32}$ years.}
\label{IND1}
\end{figure}
%---------------------------------------------------------

\begin{figure}[ttt]
\begin{center}
\includegraphics[scale=0.6]{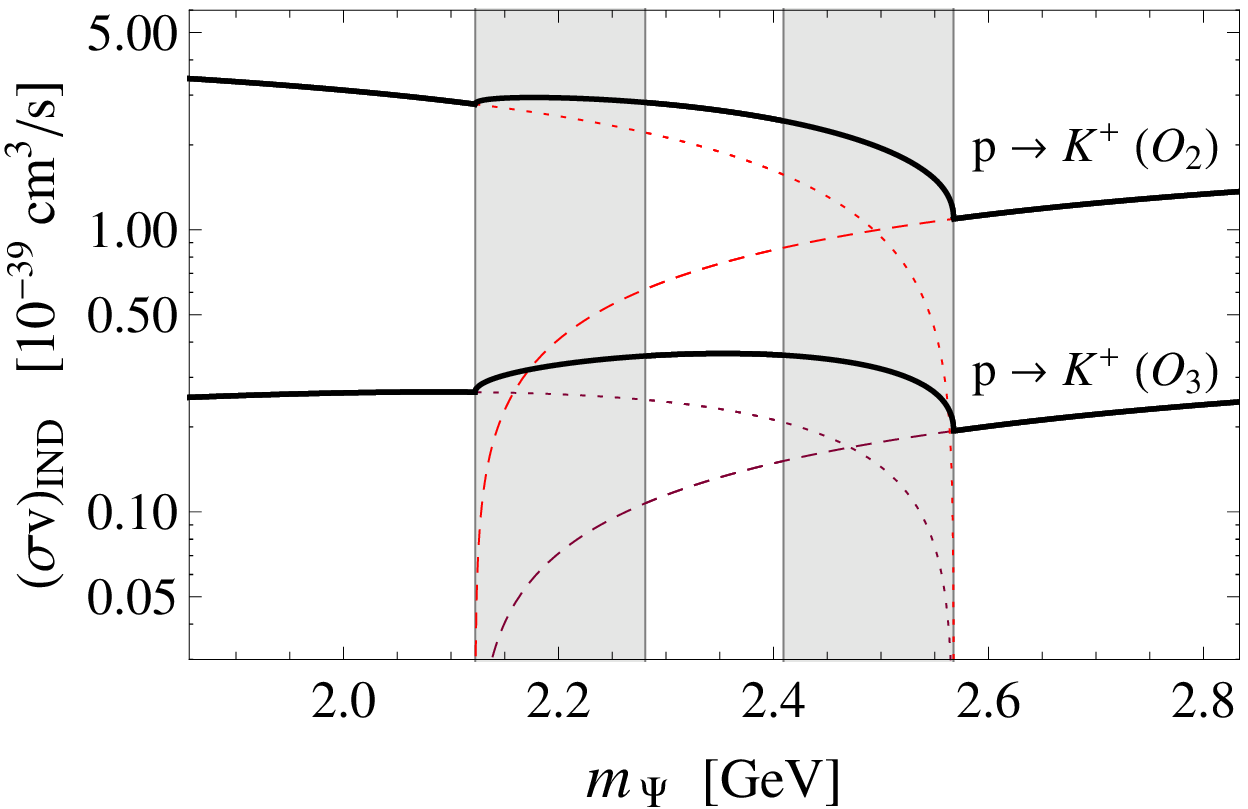} \quad \includegraphics[scale=0.6]{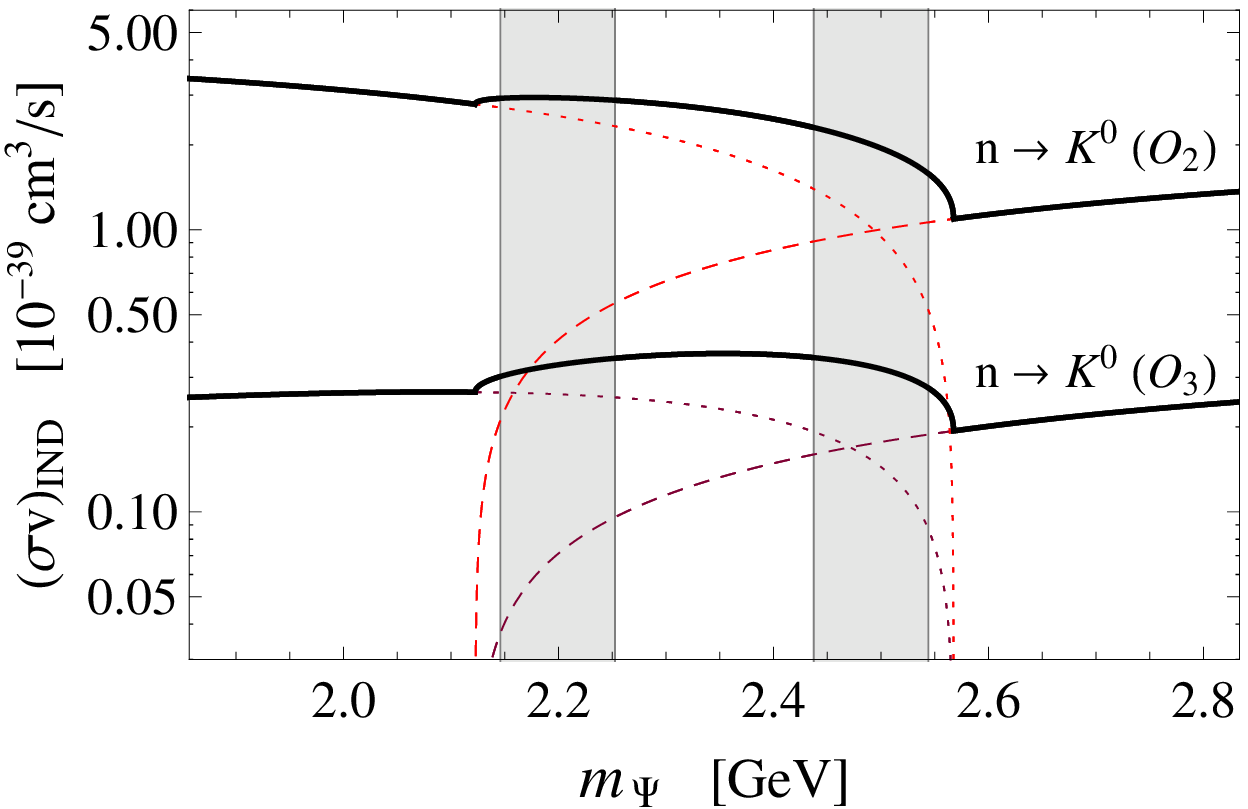}
\end{center}
\vspace{-.4cm}
\caption{\it \small Induced nucleon decay cross
sections $(\sigma v)_{IND}$ for $p \to K^+$ (left) and $n \to K^0$
(right) as a function of fermion DM mass $m_{\Psi}$ for $|c_{2,3}|
=\textrm{TeV}^{-3}$. Dotted (dashed) lines denote $N \Phi \to
\bar{\Psi} K$ ($N \Psi \to \Phi^\dagger K$) from operators
$O_{2,3}$.  Solid lines denote total rates $N \Phi \to \bar{\Psi} K$
+ $N \Psi \to \Phi^\dagger K$.  Grey regions show where existing
nucleon decay bounds apply, described in text.}
\label{IND2}
\end{figure}

Our numerical results are shown in Figs.~\ref{IND1} and \ref{IND2}, for the five decay modes we consider.  We plot the velocity-weighted cross sections $(\sigma v)_{IND}$ for each mode, as a function of $m_\Psi \approx (5 m_p - m_{\Phi})$, for the allowed range $2m_p < m_\Psi < 3 m_p$ (as required by DM stability).  In Fig.~\ref{IND1}, the solid lines show the total cross sections for the channels $p \to \pi^+$ and $n \to \pi^0$ (left) and $n \to \eta$ (right).  All three modes arise from the same IND operator $O_1$, and we have fixed $c_1 = \textrm{TeV}^{-3}$.  The individual cross sections for $N \Psi \to M \Phi^\dagger$ ($N \Phi \to M \bar{\Psi}$) are shown by the dotted (dashed) curves.  Although the individual rates vanish where kinematically forbidden, the total rate is always non-vanishing.  For $|m_\Psi - m_\Phi| < m_N - m_M$, both up- and down-scattering rates are non-zero, and IND is bichromatic.  Otherwise, only down-scattering is allowed, and IND is monochromatic.

%---------------------------------------------------------
%---------------------------------------------------------

In Fig.~\ref{IND2}, we show the strange IND channels $p \to K^+$ (left) and $n \to K^0$ (right).  Here, there are two relevant operators $O_{2,3}$.  The solid lines denote the total IND cross sections induced by each $O_i$ independently, taking $c_i = \textrm{TeV}^{-3}$, for $i=2,3$.  The individual cross sections for $N \Psi \to M \Phi^\dagger$ ($N \Phi \to M \bar{\Psi}$) are shown by the dotted (dashed) curves.  Again, the kaons can be monochromatic or bichromatic, depending on whether up-scattering is kinematically allowed.

Existing nucleon decay bounds do apply in select regions of parameter space for up-scattering close to threshold, where the meson momentum is reduced.  We illustrate these regions in Fig.~\ref{IND2}, shown in grey, for the case of Super-Kamiokande.  For $p \to K^+$, this region corresponds to $\beta_{K^+} < 0.75$:
in this case, the $K^+$ is below \v{C}erenkov threshold and the event topology is identical to $p \to K^+ \nu$.  For $n \to K^0$, this region corresponds to a kinematic window $200 < p_{K^0} < 500$ MeV, as in their $n \to K^0 \nu$ search.  We emphasize that Super-Kamiokande bounds constrain only the up-scattering IND rate, which can be suppressed compared to down-scattering in these parameter regions.

The total IND rate depends sensitively on the unknown mass scale $\Lambda_{IND} \equiv |c_i|^{-1/3}$.  The total IND cross sections and nucleon lifetimes for all channels are comparable, scaling as
\beq
(\sigma v)_{IND} \approx 10^{-39} \; \textrm{cm}^3/\textrm{s} \times \left( \frac{\Lambda_{IND}}{1 \; \textrm{TeV} } \right)^{-6} \, , \quad
\tau_{N} \approx 10^{32} \; \textrm{yr} \times \left( \frac{\Lambda_{IND}}{1 \; \textrm{TeV} }
\right)^{6}\left( \frac{\rho_{DM}}{0.3 \; \textrm{GeV}/\textrm{cm}^3 } \right) \, .
\eeq
As we show in Sec.~\ref{sec:coll}, the collider bound on this scale is $\Lambda_{IND} \gtrsim 300$ GeV.
Therefore, $\tau_{N}$ can in principle be as low as $10^{29}$ years.  It is likely that such a short
lifetime would be excluded from nucleon decay searches, but no dedicated IND search has yet been performed.  We also note that
future nucleon decay experiments are envisioned to have markedly better reach for $\tau_N$.  For example,
a water \v{C}erenkov detector with $10^4$ kton $\times$ year of exposure
can reach $\tau_N \sim 10^{(34-35)}$~yr \cite{Suzuki:2001rb,Diwan:2003uw}; a similar capability is expected to be
achieved with a liquid Argon detector with $10^3$ kton $\times$ year of exposure \cite{Bueno:2007um}.  For kaon
final states, liquid Argon technology is expected to provide improved efficiency due to better
imaging capabilities \cite{Bueno:2007um}.

Lastly, we note that the IND cross sections satisfy certain relations, as a consequence of strong isospin symmetry:
\beq
(\sigma v)_{IND}^{p \to \pi^+} = 2 \, (\sigma v)_{IND}^{n \to \pi^0} \, , \quad (\sigma v)_{IND}^{p \to K^+} = (\sigma v)_{IND}^{n \to K^0} \; .
\eeq
The latter relation holds only if $N \to K$ modes are dominated by either $O_2$ or $O_3$, as assumed in Fig.~\ref{IND2}.
If neither operator is negligible, then $(\sigma v)_{IND}^{p \to K^+} \ne (\sigma v)_{IND}^{n \to K^0}$.
In this case, both $K$ modes are complementary and can be used to disentangle the underlying IND operator structure.

\section{Collider Signals from Hylogenesis\label{sec:coll}}

  ADM scenarios rely on transfer operators to connect global charge
between the visible and dark sectors, and these operators can be
probed at high-energy colliders.
In the specific hylogenesis model of Ref.~\cite{Davoudiasl:2010am},
$B$ is mediated between the two sectors by heavy Dirac fermions
$X_{1,2}$ ($m_{X_1} < m_{X_2}$), through interactions of the form
\beq
-\mathscr{L} \supset
\sum_{a=1,2} \frac{\lambda_a^{ijk}}{M^2}\,
(u_R^i d_R^j)({X}_{a,L}^{\dagger} d^k_R)
+ \zeta_a \, ({X}_{a,L}\Psi_L+X_{a,R}\Psi_R) \Phi + \textrm{h.c.}
\label{eq:couplings}
\eeq
where $i,j,k$ label generation, color indices are implicitly
contracted antisymmetrically,
and other fermion contractions are also possible.
Integrating out $X_{1,2}$ generates operators of the form of Eq.~\eqref{INDop}.

  The $X_{1,2}$ particles can be produced at high-energy hadron
colliders through the operator of Eq.~\eqref{eq:couplings}.
With decays $X_{1,2} \to \bar{\Psi} \Phi^\dagger$, this gives rise to events
involving missing energy and one or more jets.  In this section,
we investigate the sensitivity of the Tevatron and the LHC to such
events and we derive a corresponding bound on the heavy mass scale $M$
suppressing the neutron portal operator.  Related studies in the context of
WIMP and other dark matter candidates can be found in
Refs.~\cite{Birkedal:2004xn,Beltran:2010ww,Goodman:2010yf,
Bai:2010hh,Fox:2011fx,Buckley:2011kk}.

  To be concrete, we will focus on the lighter state $X_1 \equiv X$, with
the specific interaction
\bea
- \mathscr{L} &\supset&  \frac{\lambda}{M^2}
({X}^{\dagger}_L s_R)({u}_R d_R) + \zeta\,X\Psi\Phi + h.c.,
\label{op1}
\eeq
We expect other flavour structures and fermion contractions to
give qualitatively similar results.
The operator of Eq.~\eqref{op1} can give rise to processes of the form
\bea
q(p_1)\, q^\prime(p_2) \, \to \, \bar{q}^{\prime\prime}(p_3)\, \bar{\Psi}(p_4)\, \Phi^\dagger(p_5)
\eea
through either
a real or off-shell $X$, where $q,q',q''=u,d,s$ quarks.
The corresponding summed and averaged squared
matrix element can take two possible forms, depending on how the fermions
are contracted.  They are:
\beq
{|\mathcal{M}|^2} = \left\{
\begin{array}{cc}
\frac{2}{3}\left|\frac{\lambda\,\zeta}{M^2}\right|^2
\left|\frac{1}{q^2-m_x^2+i\Gamma_xm_x}\right|^2
(p_1\cdot p_2)
\left[2(p_3\ccdot q)(p_4\ccdot q)-(q^2-m_X^2)(p_3\ccdot p_4)\right]~;&
~~~~~s\text{-like}\\
&\\
\frac{2}{3}\left|\frac{\lambda\,\zeta}{M^2}\right|^2
\left|\frac{m_x}{q^2-m_x^2+i\Gamma_xm_x}\right|^2
(p_1\cdot p_3)
\left[2(p_2\ccdot q)(p_4\ccdot q)-(q^2-m_X^2)(p_2\ccdot p_4)\right]~;&
~~~~~t\text{-like}
\end{array}\right.
\label{matrixel}
\eeq
Here, $q = (p_4+p_5) = (p_1+p_2-p_3)$ is the momentum carried by
the intermediate $X$ state, and $\Gamma_x = \zeta^2m_X/16\pi$ is the
width of the $X$ state which we assume decays mainly into $\bar{\Psi}\Phi^\dagger$.
The $s$-like form correponds to the case where both initial state
fermions are contracted together in the underlying operator
while the $t$-like form corresponds to a contraction between
initial and final state quarks.

  At the Tevatron and the LHC, we find that the full cross section
derived from the matrix elements of Eq.~\eqref{matrixel} is
frequently dominated by the pole in the intermediate $X$ propagator.
This corresponds to the production of an on-shell $X$ state whose mass
is not much smaller than the higher-dimensional operator scale $M$,
and corresponds to a large momentum transfer.\footnote{
This is a necessary condition for hylogenesis to create
a sufficiently large asymmetry.}  In this case the higher-dimensional
operator structures we are using become unreliable and the full
dynamics of the unknown ultraviolet completion becomes relevant.
Without specifying the underlying theory, we can still
parametrize the generic behaviour in a reasonable way as follows.
For $s$-like contractions, whose structure would arise most naturally
from a boson in the $s$-channel, we make the replacement
\beq
\frac{\lambda}{M^2} \to \frac{\lambda}{\hat{s}-M^2+i\sqrt{\hat{s}}\Gamma},
\eeq
where $\hat{s}$ is the parton-level Mandelstam variable and $\Gamma$
is the decay width of the mediator.  In the absence of an underlying
theory, we parametrize this quantity as $\Gamma = \mathcal{C}M$
and consider the values $\mathcal{C} = 1/5,\,1/50$.
For $t$-like contractions, we make the replacement
\beq
\frac{\lambda}{M^2} \to \frac{\lambda}{\hat{t}-M^2},
\eeq
where $\hat{t} = (p_1-p_3)^2$ is the parton-level Mandelstam variable.
Since $\hat{t}$ is negative, we do not bother adding a width term.

%---------------------------------------------------------
\begin{figure}[ttt]
\begin{center}
        \includegraphics[width = 0.75\textwidth]{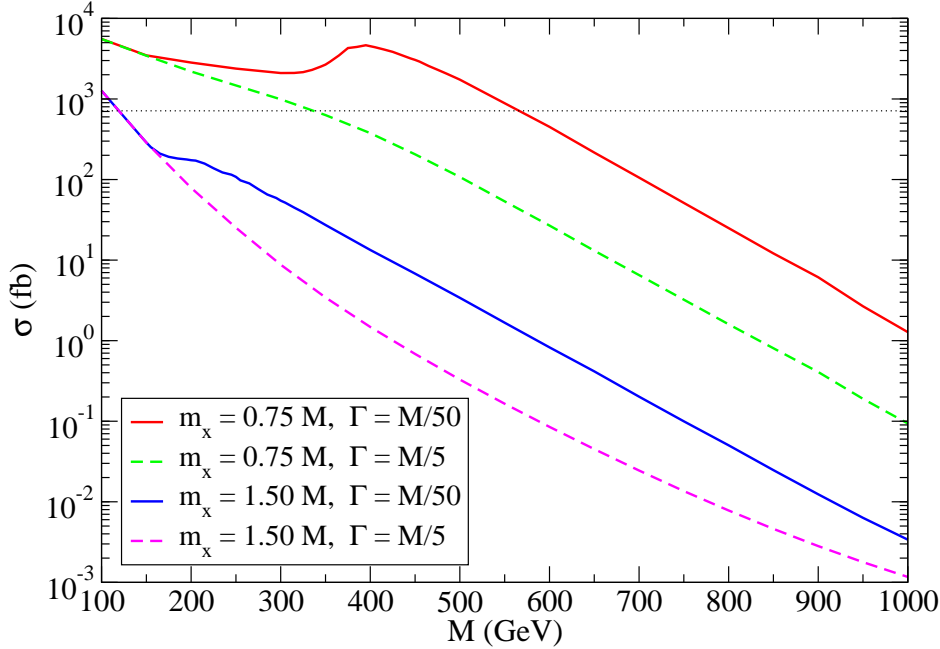}
\end{center}
\caption{Leading-order monojet production cross sections at the
Tevatron subject to the cuts described in the text.  We show lines for
$m_X = 0.7\,5M,\,1.50\,M$ and $\Gamma = M/5,\,M/50$, and we set
$\lambda = 1$ and $\zeta = 0.7$.}
\label{tevatron}
\end{figure}
%---------------------------------------------------------

  We apply these matrix elements to compute the leading-order (LO)
monojet production cross sections at the Tevatron by convolving with
CTEQ6.1M parton distribution functions~\cite{Pumplin:2002vw}
and integrating over phase space.
To match the most stringent Tevatron monojet
search bounds, we impose a cut of $p_T > 80\,\gev$ and $|\eta| < 1.0$ on
the outgoing jet.  Following Ref.~\cite{Goodman:2010yf}, we also apply
a flat efficiency factor of $40\%$ to connect our parton-level
cross section to the full hadronic jet reconstruction at
the Tevatron detectors.  The cross sections computed in this way
(after applying cuts and the efficiency factor) are shown
in Fig.~\ref{tevatron} for the parameter values
$m_X = 0.75M,\,1.5M$ and
$\Gamma = M/5,\,M/50$.  We also set $\lambda = 1$ and $\zeta = 0.7$.
This figure shows a significant resonant enhancement when the intermediate
state is narrow.  A similar enhancement was seen in Ref.~\cite{Fox:2011fx}.
The resonant enhancement only becomes fully operational at
$M \gtrsim 300\,\gev$ due to the cut on jet $p_T$.
The horizontal dashed line in Fig.~\ref{tevatron}
represents the current Tevatron $2\sigma$ exclusion limit on the net monojet
cross section of $664\,fb$ (after cuts)~\cite{Goodman:2010yf},
based on the CDF analysis of Refs.~\cite{Aaltonen:2008hh,cdfmono} which
uses the same set of jet cuts as applied to our signal estimates.
This limit translates into a lower bound of $M = 200\!-\!700\,\gev$
for the operator of Eq.~\eqref{op1}, depending on the mass of the
$X$ state and the width of the unspecified intermediate state.

\begin{figure}[ttt]
\begin{center}
        \includegraphics[width = 0.75\textwidth]{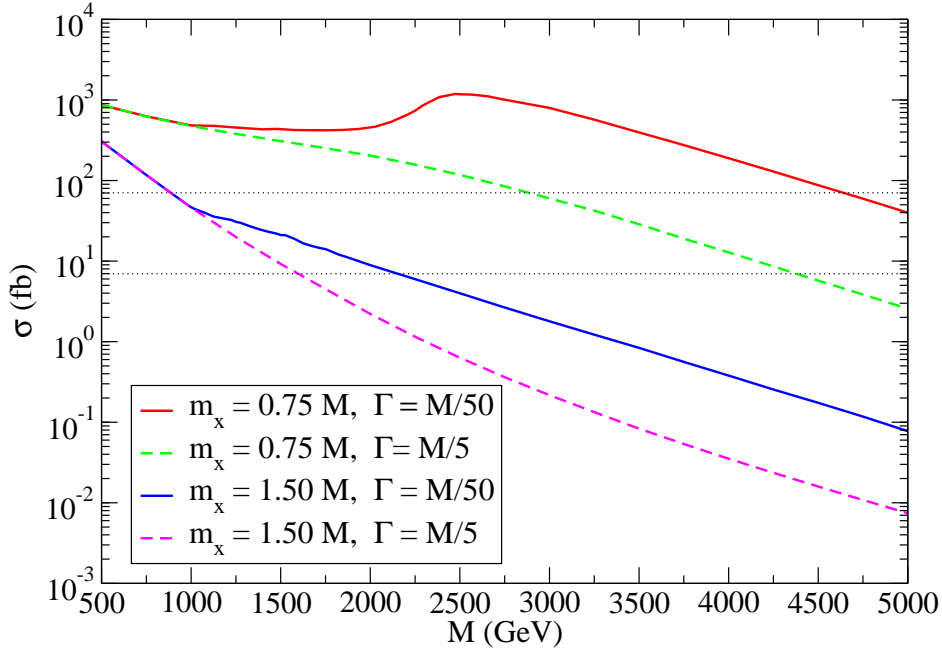}
\end{center}
\caption{Jet plus missing energy production cross sections
at the LHC ($14\,\tev$) subject to the cuts described in the text.
We show lines for $m_X = 0.75\,M,\,1.50\,M$ and $\Gamma = M/5,\,M/50$,
and we set $\lambda = 1$ and $\zeta = 0.7$.}
\label{lhc}
\end{figure}

  Monojet signals can also be detected at the LHC.  Due to the expectation of
a significant amount of associated QCD radiation, Ref.~\cite{Vacavant:2001sd}
investigated the reach of an inclusive search for a hard jet plus missing
energy search at ATLAS with no veto on additional hard jets.
To match this analysis, we compute the inclusive leading-order
parton-level cross section at the LHC with $\sqrt{s}=14\,\tev$
subject to the cuts $p_T > 500\,\gev$ and $|\eta| < 3.2$ on the outgoing jet.
We also rescale the cross section by a conservative acceptance/efficiency
factor of $85\%$~\cite{Goodman:2010yf}.  The corresponding cross sections
are shown in Fig.~\ref{lhc} for several values of $M$ with $m_X = 0.75M,\,1.5M$
and $\Gamma = M/5,\,M/50$.  We also set $\lambda =1$ and $\zeta = 0.7$.
Based on the background analysis of Ref.~\cite{Vacavant:2001sd},
the study in Ref.~\cite{Goodman:2010yf} estimated a net SM background
production rate after the applied cuts of $\sigma_{BG} \simeq 200\,fb$.
Applying a simple $S/\sqrt{B}>5$ measure on the detection significance,
this leads to a sensitivity to monojet cross sections as small as
$70\,fb$ ($7\,fb$) with $1\,fb^{-1}$ ($100\,fb^{-1}$) of data
at $\sqrt{s} = 14\,\tev$.  This is shown by the dotted lines in Fig.~\ref{lhc}.
The resulting LHC reach for the operator of Eq.~\eqref{op1}
lies in the range $M=1\!-\!4\,\tev$.  This is competitive with the
sensitivity to this operator from searches for IND in existing
nucleon decay experiments.

%---------------------------------------------------------
%---------------------------------------------------------

  Additional quark operators beyond that given in Eq.~\eqref{op1} and
considered above are expected to yield qualitatively similar collider rates
and signatures in most cases.  An interesting further possibility are
quark operators involving top or bottom quarks.  With a bottom quark in
the final state, the monojet signal could be augmented with a $b$-tag.
In the case of a top quark in the final state, the signal would be
a hard single top quark with large missing energy.  While the
search reach in monobottom and monotop channels is likely greater
than for the light quark channels, the operators involving 
third-generation quarks do not correlate directly with IND processes.

The hylogenesis model in Ref.~\cite{Davoudiasl:2010am} also contains
a hidden $U(1)'$ gauge symmetry that couples to $(\Psi,\Phi)$,
is spontaneously broken at the GeV-scale, and couples to the SM only
through kinetic mixing with hypercharge.
Direct pair production of DM particles via the corresponding $Z'$
vector boson with an associated jet can be another source of
monojet signals~\cite{Goodman:2010yf,Bai:2010hh}.
However, since this vector is relatively light, the existing bounds from the
Tevatron and the expected reach of the LHC are both much weaker than
the limits from DM direct detection through elastic scattering mediated
by the $Z'$.

\section{IND and Stellar Evolution
\label{sec:star}}

  DM can be captured in stars by elastic scattering with the nuclei
they contain.  Once captured, ordinary self-annihilating DM will collect
in the middle of the star and annihilate with other DM particles there,
releasing energy~\cite{Press:1985ug,Griest:1986yu,Gould:1987ir}.
Self-annihilation is not possible for hylogenic dark matter~(hDM)
due to its conserved global charge.  Instead, it can potentially destroy
baryons within the star via IND processes such as
$\Psi \, N \to \Phi^\dagger  M$.
If it remains within the star, the anti-hDM $\Phi^{\dagger}$ reaction
product can then annihilate with a $\Phi$ particle already captured
in the stellar core.  The net result of this chain is the destruction
of a baryon and the release of energy, both from the meson decay and
the annihilation step.

  In this section we investigate the effects of IND processes
on several varieties of stellar species including neutron stars,
white dwarfs, and main-sequence stars.  To be concrete,
we will concentrate on the specific model of hylogenesis presented
in Ref.~\cite{Davoudiasl:2010am} and described in
Appendix~\ref{sec:appa} with IND mediated by the operator of Eq.~\eqref{INDop}.
We will assume a fiducial spin-independent {proton} scattering
cross section of $\sigma^{SI}_p = 10^{-39}\,\textrm{cm}^2$
(and $\sigma^{SI}_n=0$) for both $\Psi$ and $\Phi$ induced by their
coupling to the kinetically-mixed $U(1)'$ vector
boson,\footnote{For DM masses below $3\,\gev$, this value is
consistent with existing direct detection searches~\cite{Angloher:2002in,
Ahmed:2010wy,Aalseth:2010vx}.}
as well as an $hDM$-anti-$hDM$ annihilation cross section of
$(\sigma v)_{ann} = 10^{-25}\textrm{cm}^3/\textrm{s}$.  We shall consider two cases for
the IND cross section: a large value of $(\sigma v)_{IND} = 10^{-39}\textrm{cm}^3/\textrm{s}$,
and a small value of $(\sigma v)_{IND} = 0$.  The general behaviour
for other IND rates will lie somewhere between these two extremes.

\subsection{Stellar Capture and Annihilation}

  Relic $\Psi$ and $\Phi$ particles will be captured in stars by
scattering with nuclei to energies below the local escape velocity.
Once captured, an hDM particle will undergo further scatterings,
thermalize with the baryons in the star, and collect within
the stellar core.  This occurs quickly relative to the
lifetimes of the stars we consider here for our large fiducial value
of the proton scattering cross section.  Once they thermalize,
the DM particles are largely confined to a core of
radius~\cite{Griest:1986yu,Gould:1987ir,Bertone:2007ae}
\beq
r_{i,th} = \lrf{9T_c}{4\pi G\rho_cm_i}^{1/2},
\eeq
where $i = \Psi,\,\Phi$, $T_c$ is the mean temperature
and $\rho_c$ is the mean (baryon) density in the stellar core.

    The evolution of the total numbers of $\Psi$, $\Phi$,
$\bar{\Psi}$, and $\Phi^\dagger$ within a star is described the following
system of equations:
\bea
\frac{dN_{\Psi}}{dt} &=& C_\Psi
- A_{\Psi}N_{\Psi}N_{\bar{\Psi}}
- B_{\Psi}N_{\Psi}
%- E_{\Psi}N_{\Psi}
\label{neqy}\\
\frac{dN_{\bar{\Psi}}}{dt} &=& \phantom{C_\Psi}
- A_{\Psi}N_{\Psi}N_{\bar{\Psi}}
+ \epsilon_{\bar{\Psi}}B_{\Phi}N_{\Phi}
%- E_{\Psi}N_{\bar{\Psi}}
\label{neqyb}\\
\frac{dN_{\Phi}}{dt} &=& C_{\Phi} - A_{\Phi}N_{\Phi}N_{{\Phi}^\dagger}
- B_{\Phi}N_{\Phi}
%-E_{\Phi}N_{\Phi}
\label{neqf}\\
\frac{dN_{{\Phi}^\dagger}}{dt} &=&
\phantom{C_\Psi}
- A_{\Phi}N_{\Phi}N_{{\Phi}^\dagger}
+ \epsilon_{{\Phi}^\dagger}B_{\Psi}N_{\Psi}
%-E_{\Phi}N_{{\Phi}^\dagger}
\label{neqfb}
\eea
Here, the $C_i$ coefficients are the hDM capture rates,
the $A_i$ coefficients describe hDM-anti-hDM annihilation, and
the $B_i$ coefficients describe IND.
A general expression for $C_i$ is given in Ref.~\cite{Gould:1987ir},
while the $A$ and $B$ coefficients are given to a good approximation by
\bea
A_i &\simeq& (\sigma v)_{i,ann} \bigg/
\left(4\pi r_{i,th}^3/3\right),\\
B_i &\simeq& (\sigma v)_{i,IND} \;(\rho_c/m_n),
\eea
where $m_n$ is the mass of a nucleon.
The $\epsilon_i$ terms appearing in Eqs.~\eqref{neqyb} and \eqref{neqfb}
are the probabilities for the anti-hDM products of IND to be captured
by the host star after they are created.
In certain regimes additional processes can influence the evolution
of the stellar populations of hDM and anti-hDM such as
evaporation~\cite{Griest:1986yu,Gould:1987ju} and direct annihilation
to baryons (\textit{e.g.} $\Psi \, \Phi\to \bar{N} \, M$).  We will discuss
these effects when they may be relevant.

\subsection{Neutron Stars}

  Neutron stars are very dense objects supported by the Fermi
degeneracy pressure of their neutrons.  Despite their name, they also
contain a significant mass fraction %(over $10\%$)
of protons and heavier nuclei, and the nuclear state of
their cores is not fully understood~\cite{Hebeler:2010jx}.
Typical neutron star parameters are mass $M=1.4M_{\odot}$,
with $M_\odot \simeq 2.0\times 10^{30}$~kg the solar mass,
radius $R=10\,\textrm{km}$, core temperature $T_c = 10^5\,\textrm{K}$, and core
baryon density $\rho_c = 1.4\times 10^{18}\textrm{kg\,/m}^3$~\cite{
Bertone:2007ae,Goldman:1989nd,Kouvaris:2007ay,deLavallaz:2010wp}.

  The rate of capture of $\Psi$ or $\Phi$ upon a typical neutron
star (including general relativistic corrections)
is~\cite{Goldman:1989nd,Kouvaris:2007ay,deLavallaz:2010wp}
\beq
C_i \simeq 2.5\times 10^{25}\textrm{s}^{-1}\lrf{\rho_{DM}}{\gev/\textrm{cm}^3}
\lrf{5\,\gev}{m_{\Psi}+m_{\Phi}}\lrf{220\,\textrm{km/s}}{\bar{v}}\,f,
\label{nscapture}
\eeq
where $\rho_{DM}$ is the local DM energy density, $\bar{v}$ is the local
DM velocity dispersion, and
\beq
f = min\left\{1,\,(x_p\sigma_p+x_n\sigma_n)/(2\times 10^{-45} \, \textrm{cm}^2)\right\},
\eeq
with $x_p$ and $x_n$ being the proton and neutron mass fractions.
The factor $f$ accounts for the saturation of the cross section at
the cross-sectional area of the star.  This saturation sets in when the star
becomes optically thick to DM -- when a DM particle impinging upon the
star is likely to scatter multiple times with nucleons in the star.
For the fiducial nucleon scattering cross section we are using and
assuming a proton mass fraction of $x_p = 0.1$, we find that neutron stars are
optically thick to both $\Psi$ and $\Phi$.  This leads to
$C_{\Psi}=C_{\Phi}\equiv C$ as well as $\epsilon_{\Psi}=1=\epsilon_{\Phi}$.
We also find
\beq
r_{i,th}
%= \lrf{9T_c}{4\pi G\rho_Bm_i}^{1/2}
\simeq (140\,\textrm{cm})\lrf{T_c}{10^5\,\textrm{K}}^{1/2}\lrf{3\gev}{m_i}^{1/2}
\lrf{1.4\times 10^{18}\textrm{kg/m}^3}{\rho_c}^{1/2},
\eeq
implying $A_i \sim 5\times 10^{-32}\textrm{s}^{-1}$ 
and $B_i\sim 0.9\,\textrm{s}^{-1}\,(0\,\textrm{s}^{-1})$
for the large (small) IND rate 
$(\sigma v)_{IND} = 10^{-39}\textrm{cm}^3/\textrm{s}$ 
($0 \, \textrm{cm}^3/\textrm{s}$).
%Evaporation is negligible for neutron stars and we set
%$E_i \to 0$~\cite{McDermott:2011jp}.

\begin{figure}[ttt]
\begin{center}
        \includegraphics[width = 0.75\textwidth]{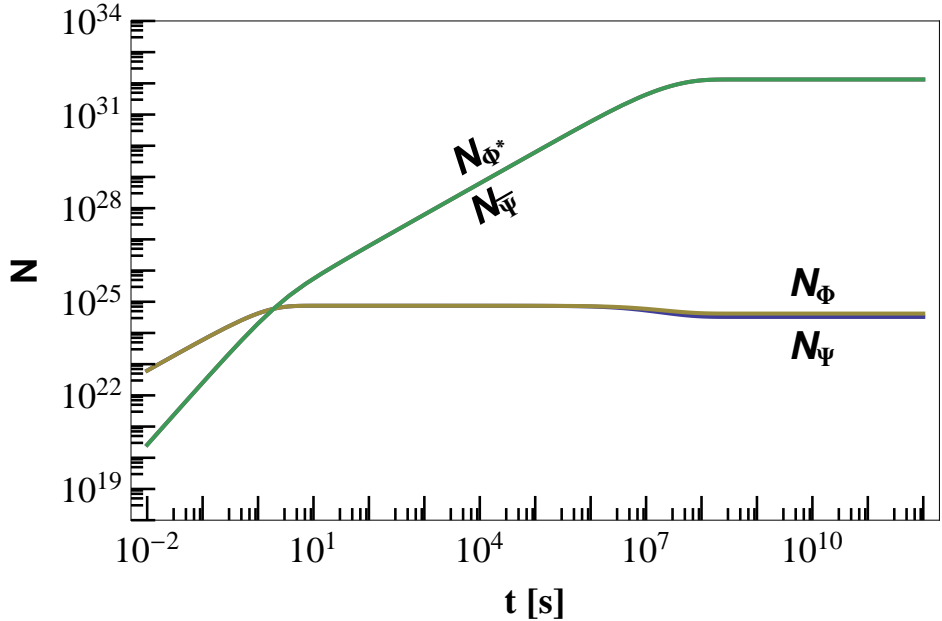}
\end{center}
\caption{The build-up of the abundance of hDM and anti-hDM in a neutron star for $m_{\Psi}=2.85~{\rm GeV}$ and $m_{\Phi}=2.05~{\rm GeV}$ for a total dark matter density of $0.3~{\rm GeV/\textrm{cm}^{3}}$.
At this density a steady state is attained after 
a little over $10^7\,\textrm{s}$.}
\label{neutron}
\end{figure}

  Consider first the case of a large IND rate.
Using the evolution equations of Eqs.~(\ref{neqy}--\ref{neqfb})
and assuming negligible initial hDM densities, we find that the numbers of
hDM and anti-hDM particles within a typical neutron star reach a steady state.
This behaviour is shown in Fig.~\ref{neutron} for the fiducial
cross sections given above and $m_{\Psi} = 2.85\,\gev$, $m_{\Phi} = 2.05\,\gev$.
The steady-state populations in the limit of $B_i \gg A_i$
are approximated well by the analytic expressions
\beq
N_{\Psi} \simeq \frac{C}{B_{\Psi}(1+\xi)},~
N_{\bar{\Psi}} \simeq \frac{B_{\Psi}}{A_{\Psi}}\lrf{1+\xi}{1+\xi^{-1}},~
N_{\Phi}\simeq\frac{C}{B_{\Phi}(1+\xi^{-1})},~
N_{{\Phi}^\dagger}\simeq \frac{B_{\Phi}}{A_{\Phi}}\lrf{1+\xi^{-1}}{1+\xi},
\eeq
with $\xi = A_{\Psi}B_{\Phi}/A_{\Phi}B_{\Psi}$.\footnote{
For $A_i\gg B_i$, we find:
$N_{\Psi} = N_{\Phi} = C/(B_{\Psi}+B_{\Phi}),~N_{\bar{\Psi}}=B_{\Phi}/A_{\Psi},~
N_{{\Phi}^\dagger}=B_{\Psi}/A_{\Phi}$.}
A steady state is also attained when one of the IND rates vanishes due
to kinematic suppression.  For example, with $B_{\Psi}\gg A_{\Psi,\Phi}$
and $B_{\Phi}\to 0$, we find
\beq
N_\Psi= \frac{C}{B_\Psi},~~~
N_{\bar{\Psi}}=0,~~~
N_{\Phi,{\Phi}^\dagger}= \pm \frac{C}{2B_\Psi}
+ \sqrt{\lrf{C}{2B_\Psi}^2+\frac{C}{A_{\Phi}}},
\eeq
where in the last equality the plus sign corresponds to $\Phi$ and
the minus sign to $\Phi^{\dagger}$.
The solution for $B_{\Psi}\to 0$ and $B_{\Phi}\gg A_{\Psi,\Phi}$ is
identical but with $\Phi$ and $\Psi$ interchanged in the expressions above.
In both cases, the steady-state particle populations are on the order of
$N_{\Psi,\Phi} \simeq 10^{25}$ and
$N_{\bar{\Psi},{\Phi}^\dagger} \simeq 10^{31}$
(when they are non-zero) for the fiducial input values listed above.

  The time needed to reach this steady state from a negligible
initial dark matter density in the star is on the order of
\beq
\tau_{ss} \sim max\left\{B^{-1},\frac{B}{CA}\right\}.
\eeq
where the first value corresponds to IND balancing capture and the second
to annihilation balancing production of anti-hDM by IND.  For the fiducial
cross sections we are considering, we find that annihilation
takes longer to balance, and leads to $\tau_{ss} \sim 2\times10^{7}\,\textrm{s}$.
This  is ultra short relative to the lifetime of a typical
neutron star.

%---------------------------------------------------------
%---------------------------------------------------------

  Having reached a steady state, the main combined effect of the IND
and annihilation processes is to inject energy into the host neutron
star with rate $(m_\Psi+m_{\Phi}+m_N)C$.  In this respect, hDM in the steady
state limit has the same effect on neutron stars as ordinary
self-annihilating DM.  Energy injection by DM can interfere
with and halt the cooling of old neutron stars, and therefore the observation
of a very cool, old neutron star in a region of large DM density could
put significant constraints on a wide variety of DM
scenarios~\cite{Bertone:2007ae,Kouvaris:2007ay,deLavallaz:2010wp}.
However, this effect is too small to be observed using existing observations,
and appears to be challenging to probe in the near
future~\cite{Bertone:2007ae,Kouvaris:2007ay,deLavallaz:2010wp}.
With hDM, baryons within the neutron star are also destroyed by IND,
but the number is negligible compared to the total of
$N_B \simeq 2 \times 10^{57}$
within a typical neutron star over the lifetime of the Universe unless
the local DM density approaches an enormous value 
of $10^{14}\gev/\textrm{cm}^3$.

  Consider next the case of a vanishingly small IND cross section.
The populations of of $\Psi$ and $\Phi$ hDM particles will now
build up within the neutron star with rate given by Eq.~\eqref{nscapture},
corresponding to populations of about $10^{43}(\rho_{DM}/\gev\,\textrm{cm}^{-3})$
over the lifetime of the Universe.  We can compare this number
to the populations required for DM particles to begin self-gravitating
and to form a black hole.  Self-gravitation begins
when~\cite{Bertone:2007ae,Kouvaris:2010jy,McDermott:2011jp}
\beq
N_i \gtrsim N_{self}\equiv \frac{\rho_c}{m_i}(4\pi r_{i,th}^3/3)
\simeq 3\times 10^{45}\lrf{3\gev}{m_i}^{5/2}\lrf{T_c}{10^5\textrm{K}}^{3/2}
\lrf{1.4\times 10^{18}\textrm{kg/m}^3}{\rho_c}^{1/2}\!\!\!\!\!.
\label{nself}
\eeq
This is much greater than the steady-state populations found above for the case
of a large IND rate, and also greater than the numbers collected for a
smaller IND rates for local hDM densities below
$(3 \times 10^2\gev/\textrm{cm}^3)\,
min\{1,\;3\times 10^{-57}\textrm{cm}^3\textrm{s}^{-1}/(\sigma v)_{IND}\}$.

  Self-gravitating DM particles will form a black hole that can potentially
destroy the host star unless they are stabilized in
some way~\cite{Bertone:2007ae,Goldman:1989nd,Kouvaris:2010jy,McDermott:2011jp,Gould:1989gw}.
For non-interacting fermions there is a degeneracy pressure that must
be overcome.  To do so, the total number of fermions must exceed
\beq
N_i \gtrsim N_{crit}^f \equiv \lrf{\sqrt{8\pi}M_\text{Pl}}{m_i}^3
\simeq 6\times 10^{55}\lrf{3\,\gev}{m_i}^3,
\label{ncritf}
\eeq
with $M_\text{Pl} = \sqrt{8\pi/G}\simeq 2.4\times 10^{18}\gev$ the reduced
Planck mass.
In the case of non-self-interacting bosons, there is still a zero-point
pressure which can be overcome if the number of bosons exceeds
\beq
N_i \gtrsim N_{crit}^{b} \equiv \lrf{\sqrt{8\pi}M_\text{Pl}}{m_i}^2
\simeq 2\times 10^{37}\lrf{3\,\gev}{m_i}^2.
\eeq
A bosonic black hole can also arise even before bulk self-gravitation sets
in through the formation of a Bose-Einstein
condensate~\cite{Kouvaris:2010jy,McDermott:2011jp}.

  We argue that black holes are not likely to form in the specific
theory of hylogenesis discussed in Ref.~\cite{Davoudiasl:2010am}
and Appendix~\ref{sec:appa}, in which $\Phi$ and $\Psi$ carry
equal and opposite
charges under a spontaneously broken $U(1)'$ gauge symmetry,
until $N_{\Psi,\Phi} > N_{crit}^f$.
At distances much smaller than the inverse vector mass $m_{Z'}^{-1}$, the effects
of breaking the $U(1)'$ can be neglected and the $Z'$ vector boson mediates
a repulsive force between particles with like-sign charges.  This induces
a pressure among $\Phi$ particles that prevents them from collapsing into
a black hole provided $m_{Z'} \ll (m_{\Phi}M_\text{Pl}^2)^{1/3}$
and $e' \gg m_{\Phi}/M_\text{Pl}$ (where $e'$ is the $U(1)'$ coupling
of $\Phi$)~\cite{Kouvaris:2010jy}.
Instead, under these mild assumptions the formation of a black
hole requires an approximately charge-neutral collection of hDM particles
and therefore roughly equal numbers of $\Psi$ and $\Phi$ states.
A necessary condition for this to occur is $N > N_{crit}^f$ to overcome the
fermion degeneracy pressure.  Therefore we do not expect this specific
theory of hylogenesis to lead to the formation of hDM black holes
within neutron stars unless the local hDM density around the host
star exceeds $5\times 10^{11}\,\gev/\textrm{cm}^3$.

  Between the two extremes of the large and small IND rates
considered above, no new obvious observational bounds arise.
With a small but non-zero rate for IND, the $\Psi$ and $\Phi$ populations
may grow large enough to form some amount of anti-hDM that will subsequently
annihilate away.  Large hDM densities can also lead to direct annihilation of
hDM to antibaryons, $\Psi\,\Phi \to \bar{N}\,M$, and similarly for anti-hDM.
This can be accounted for by adding terms
of the form $-DN_{\Psi}N_{\Phi}$ to the evolution equations for
$\Psi$ and $\Phi$, with $D \simeq (\sigma v)_{IND}/V_{hDM}$ and
$V_{hDM}$ is the volume occupied by the hDM, whose evolution we
must also keep track of.  Note that $V_{hDM}$ appears here rather
than $4\pi r_{th}^3/3$ because this effect starts to compete with
standard IND only after the hDM population begins to self-gravitate,
when $N_i> \rho_cV_{hDM}/V > N_{self}$.  We defer an analysis of the effects
of self-gravitation and self-annihilation of hDM on the structure of
neutron stars to a future work.

\subsection{White Dwarfs}

  We can perform a similar analysis for the effects of hDM
on white dwarfs.  A typical white dwarf consists primarily of carbon and
oxygen and is supported by the degeneracy pressure of the electrons
it contains.  Typical white dwarf parameters are $R = 0.01\,R_{\odot}$,
$M = 0.7\,M_{\odot}$, $\rho_c = 10^9\,\textrm{kg/m}^3$,
$T_c = 10^7\,\textrm{K}$~\cite{Kepler:2006ns}.
We will approximate the internal structure of a white dwarf as
consisting entirely of carbon with a uniform density~\cite{Bertone:2007ae}.

  The capture rate of DM through elastic scattering with nuclei in a
white dwarf is approximated well by the expression~\cite{
Gould:1987ir,Bertone:2007ae,Bottino:2002pd,McCullough:2010ai,Hooper:2010es}
\beq
C_i \simeq \lrf{8}{3\pi}^{1/2}\!
\lrf{\rho_i\bar{v}}{m_i}\lrf{3v_{esc}^2}{2\bar{v}^2}\sigma_{eff},
\eeq
with $\bar{v}$ being the local velocity dispersion, $v_{esc}$ the escape
velocity at the surface of the star, and the the effective cross section
is defined to be
\beq
\sigma_{eff} \equiv min\left\{\pi R^2,~~
\sigma^{SI}_p\sum_k\frac{M}{m_n}\frac{x_k}{A_k}
\frac{[f_pZ_k+(A_k-Z_k)f_n]^2}{f_p^2}\lrf{m_{r_k}}{m_{r_p}}^2
\right\},
\eeq
where the sum runs over all nuclear species $k$ in the star with mass
fraction $x_k$ and atomic mass (number) $A_k$ ($Z_k$),
$\sigma^{SI}_p$ is the spin-independent nucleon scattering
cross section off protons, $f_p$ and $f_n$ are the relative
coupling strengths of the DM to protons and neutrons, and
$m_{r_k}$ ($m_{r_p}$) is the reduced mass of the DM-nucleus
(DM-proton) system.  The saturation of the effective cross section
at $\pi R^2$ corresponds to the star becoming optically thick.

  We find that a typical white dwarf is indeed optically thick for the
fiducial scattering cross section of $\sigma^{SI}_p = 10^{-39}\textrm{cm}^2$
(with $f_p=1$, $f_n=0$).  This gives
\bea
C_{\Psi}&=&C_{\Phi} \equiv C\\
&\simeq&
(6\times 10^{27}\textrm{s}^{-1})\lrf{R}{0.01R_{\odot}}\lrf{M}{0.7M_{\odot}}\!
\lrf{\rho_{hDM}}{\gev/\textrm{cm}^3}\!\lrf{5\,\gev}{m_{\Psi}+m_{\Phi}}\!
\lrf{270\,\textrm{km/s}}{\bar{v}},\nnmb
\eea
where $R_{\odot} \simeq 7.0\times 10^5$~km is the solar radius.
Once captured, hDM particles thermalize rapidly and collect within
a thermal radius at the core of the star of size
\beq
r_{i,th} \simeq (5\times 10^7\,\textrm{cm})\,\lrf{3\,\gev}{m_i}^{1/2}
\lrf{T_c}{10^7\textrm{K}}^{1/2}
\lrf{10^9\textrm{kg/m}^3}{\rho_c}^{1/2}.
\eeq
For $m_{\Psi} = m_{\Phi} = 2.25\,\gev$, we find 
$A_i = 1\times 10^{-49}\textrm{s}^{-1}$
and $B_i = 6\times 10^{-10}\textrm{s}^{-1}$ ($0\,\textrm{s}^{-1}$) 
for the case of a large (small) IND rate 
$(\sigma v)_{IND} = 10^{-39}\textrm{cm}^3/\textrm{s}$ 
($0 \, \textrm{cm}^3/\textrm{s}$)

  In the case of a large IND rate, we find that the populations
of $\Psi$, $\Phi$, $\bar{\Psi}$, and ${\Phi}^\dagger$ all reach a steady
state quickly relative to the lifetime of a typical white dwarf.
In this state we find $N_{\Psi}\simeq N_{\Phi} \simeq 5\times 10^{36}$
and $N_{\bar{\Psi}}\simeq N_{{\Phi}^\dagger} \simeq 6\times 10^{49}$ for the fiducial
parameter values listed above.
The number of baryons destroyed by IND over the age of the Universe
is a small fraction of the total number in a white dwarf provided
$\rho_{DM} \ll 10^{11}\,\gev/\textrm{cm}^3$.  The main effect of hDM capture,
IND, and annihilation on white dwarf is therefore the injection of energy into
the stellar interior with rate $(m_\Psi+m_{\Phi}+m_N)C$.  This is equivalent
to the effect of ordinary self-annihilating DM (with slightly larger mass).

  Heating of white dwarfs by DM capture and annihilation was studied
recently in Refs.~\cite{Bertone:2007ae,McCullough:2010ai,Hooper:2010es}.
These authors differ in their conclusions, with Ref.~\cite{McCullough:2010ai}
finding an upper bound on the nuclear capture cross section of about
$10^{-43}\textrm{cm}^2$ and Ref.~\cite{Hooper:2010es} finding essentially no bounds
from current observations.  The origin of the disagreement is the density
of DM within globular clusters, from which Ref.~\cite{McCullough:2010ai}
obtains their most stringent bounds, while Ref.~\cite{Hooper:2010es}
argues that they contain a much smaller dark matter abundance.
The observation of a cool white dwarf within a dwarf spheroidal
galaxy could help to resolve this question decisively~\cite{Hooper:2010es}.
Based on this uncertainty, we do not consider hDM models with
nuclear scattering cross sections as large as $\sigma_p^{SI} = 10^{-39}\textrm{cm}^2$
to be ruled out.  However, we do note that the model of hDM considered
here can also accommodate much smaller elastic scattering cross sections.

  With a small IND rate, hDM collects and builds up within the stellar
interior.  As we argued above, we do not expect any significant effects
on the star until at least $N_i \sim N_{self},\,N_{crit}^f$ given in
Eqs.~(\ref{nself},\ref{ncritf}).
For this to occur over the lifetime of the Universe, a local hDM
density approaching $10^8 -10^{10}\gev/\textrm{cm}^3$ is needed.
We do not know of any observations of white dwarfs that approach
these requirements.

\subsection{The Sun and other Main-Sequence Stars}

  The capture rate of light ($1\!-\!5\,\gev$) DM in the sun is
dominated by scattering with hydrogen and helium and is given
to a good approximation by~\cite{Gould:1987ir,Hooper:2008cf}
\bea
C_i &\simeq& (8\times 10^{25}\textrm{s}^{-1})\,
\lrf{5\gev}{m_{\Psi}+m_{\Phi}}\lrf{\rho_{DM}}{0.3\gev/\textrm{cm}^3}
\lrf{270\,\textrm{km/s}}{\bar{v}}\lrf{\sigma_p^{SI}}{10^{-39}\textrm{cm}^2}
\nnmb\\
&&~~~~~~~~~
\times \left[x_H + (1.1){x_{He}}(1+f_n/f_p)^2\frac{m_{r_{He}}^2}{m_{r_{p}}^2}\right],
\eeq
where the relative factor for helium relative to hydrogen comes from
its slightly different distribution within the sun.
In writing this expression we have also made use of the fact that the
sun is optically thin for our fiducial nucleon elastic scattering
cross section.
%*%
A second consequence of the sun being optically thin
is that nearly all the anti-hDM produced by IND escapes from the sun,
since it is typically produced with velocities much larger than $v_{esc}$.
Thus we set $\epsilon_{\bar{\Psi},\Phi^{\dagger}} \to 0$ in our evolution
equations.
%*%
The thermal radius in which the DM collects is
\beq
r_{i,th} \simeq (5\times 10^9\,\textrm{cm}) \lrf{3\,\gev}{m_i}^{1/2}
\lrf{T_c}{1.5\times 10^7\textrm{K}}^{1/2}\lrf{1.5\times 10^5\textrm{kg/m}^3}{\rho_c}^{1/2}
\eeq
which yields
%$A_i \sim 10^{-55}\,\textrm{s}^{-1}$ and
$B_i\sim 10^{-13}\,\textrm{s}^{-1}\,(0\,\textrm{s}^{-1})$ 
for the large (small) IND rate.

  An additional effect that is important for the sun is evaporation,
in which captured hDM particles are up-scattered above the escape
velocity and leave the sun~\cite{Griest:1986yu,Gould:1987ju}.
Fitting to the results of Refs.~\cite{Gould:1987ju,Hooper:2008cf},
we take the evaporation rate to be
\beq
E_i \simeq 10^{[-3.5(m_i/\gev)-4]}
\lrf{\sigma_p^{SI}}{5\times 10^{-39}\textrm{cm}^2}
\textrm{s}^{-1}.
\eeq
This evaporation adds a dissipative $-E_i N_i$ term to each
of Eqs.~(\ref{neqy}--\ref{neqfb}).  For the fiducial parameters we
are considering (and a large IND rate), evaporation becomes more important
than IND for hDM masses below $m_i\lesssim 2.4\,\gev$.

  We find that capture, evaporation, and IND reach a steady state
on a timescale of $10^{-5}$-$10^{-1}$ billion years, much shorter
than the age of the sun.  The equilibrium populations are always
less than $N_{i} \simeq C_i/max\{B_i,E_i\} \lesssim 10^{41}$,
for both large and small IND rates.
This is a very small fraction of the total mass of the sun,
and is far below what is needed for hDM to self-gravitate
or to contribute significantly to the properties of the solar
interior~\cite{Frandsen:2010yj}.  Note as well that due to the
low masses of hDM, the energies of any neutrinos produced by IND
are well below the thresholds of DM searches for upward-going muons
in neutrino telescopes such as Ice Cube~\cite{Hooper:2008cf,Abbasi:2009uz}.
The effects of hDM on other main-sequence stars are also expected to be small,
unless the star is immersed in a region of very large
DM density~\cite{Fairbairn:2007bn}.
%*%

\section{Conclusions\label{sec:conc}}

We have investigated signals of nucleon
destruction that can arise in specific theories where dark matter
is antibaryonic.  Antibaryonic DM is motivated by unified mechanisms
for DM and baryon generation (referred to as hylogenesis) which address
the cosmic coincidence between the energy densities of
dark and visible matter in the Universe.

In the hylogenesis scenario considered here,
the DM consists of an asymmetric density
of fermions $\Psi$ and scalars $\Phi$.  These particles can scatter
inelastically with nucleons via the reactions $\Psi \, N \to \Phi^\dagger \, M$
and $\Phi \, N \to \bar{\Psi} \, M$, where $N$ is a nucleon and $M$ is a meson.
These induced nucleon decay (IND) processes lead to distinctive signatures in nucleon
decay searches, stellar evolution, and hadron colliders.

IND is a novel signature of DM that can be searched for in terrestrial nucleon decay
searches.  The effective nucleon lifetime is expected to be $\sim 10^{29}-10^{32}$ years,
if baryon transfer between dark and visible sectors is mediated by new physics at a scale
$\Lambda \sim 300 \gev - 1 \tev$.
Since the DM states are unobserved, IND events mimic standard nucleon
decay with neutrino final states, but typically with greater final state meson
energy $E \sim \gev$.  Due to these different kinematics,
existing searches do not directly apply in general.
Our study therefore motivates new searches in these experiments.
We expect that the resulting sensitivities should be comparable to those in
standard nucleon decay searches.

The coupling of hylogenic DM to quarks responsible for IND can also give rise
to observable signals in hadron colliders.  In particular, such couplings
can potentially lead to monojet signals at the Tevatron and the LHC.
Existing monojet searches at the Tevatron place a lower bound on the
coupling of hylogenic DM to quarks, and this bound will be significantly
improved at the LHC.  We find that the coupling strengths that can be
probed at the LHC are of the same size as those that can produce an observable
effect in nucleon decay search experiments.  This correlation may permit
the characterization of hylogenic DM through a very diverse set of
experimental probes.

If hylogenic DM also scatters elastically with nucleons,
it can be captured in stars.  Once captured, it will thermalize with
the baryons in the star and sink to the stellar core.  In compact stars,
the local baryon density in the stellar core may be large enough 
for IND to occur at a significant rate, destroying both a nuclon 
and the hDM particle, and producing an anti-hDM particle and a meson.  
This process deposits energy in the star, an effect that may be probed 
by observing old and cool neutron stars or white dwarfs.  
However, current observations do not provide
a definitive bound on or evidence for these processes.

\section*{ Acknowledgements}

  We thank Brian Batell, Matthew Buckley, Robert Caldwell, Alejandra Castro, 
Jim Cline, Katie Freese, Patrick Fox, Michael Graesser, Ed Kearns, 
Jennifer Kile,  John Ng, Scott Oser, Maxim Pospelov, Michael-Ramsey-Musolf, 
Jen Raaf, Adam Ritz, Pat Scott, Rishi Sharma,  Hirohisa Tanaka, 
and Kathryn Zurek for helpful discussions.
We also thank Jennifer Kile for collaboration in the early stages of this work.
DM and ST would like to thank the Weizmann Institute of Science for their
hospitality.  KS would like to thank Perimeter Institute for Theoretical Physics for their hospitality.
The work of HD is supported by the United States Department
of Energy under Grant Contract DE-AC02-98CH10886.
The work of DM, KS, and ST is supported by the National Science
and Engineering Research Council of Canada~(NSERC).

\begin{appendix}
\section{Appendix\label{sec:appa}}

  In this Appendix we review some of the important features of
the hylogenesis model presented in Ref.~\cite{Davoudiasl:2010am}.
The model consists of the SM together with a SM-neutral hidden sector
containing two massive Dirac fermions $X_a$
($a=1,2$, with masses $m_{X_2} > m_{X_1} \gtrsim$ TeV),
a Dirac fermion $\Psi$, and a complex scalar $\Phi$
(with masses $m_\Psi \sim m_\Phi \sim$ GeV).
These fields couple through the ``neutron portal'' ($XU^cD^cD^c$) and
a Yukawa interaction:
\beq
-\mathscr{L} \supset
\frac{\lambda_a^{ijk}}{M^2}\,({X}_{a,L}^{\dagger}d^k_R)
(u_R^id_R^j)
+ \zeta_a \, ({X}_{a,L}\Psi_L+X_{a,R}\Psi_R) \Phi + \textrm{h.c.}
\label{eq:couplings'}
\eeq
where $i,j,k$ label flavors and the quark color indices are implicitly
contracted antisymmetrically.
Many variations on these operators exist, corresponding to different
combinations of quark flavors and spinor contractions.  With this set of
interactions one can define a conserved generalized global baryon number
with charges $B_X = -(B_\Psi + B_\Phi) = 1$.
The proton, $\Psi$, and $\Phi$ are stable due to their
$B$ and gauge charges if their masses satisfy
\beq
|m_\Psi - m_{\Phi}| < m_p + m_e,~~~~~m_p-m_e < m_\Psi + m_{\Phi} \; .
\label{eq:massdiff}
\eeq
$\Psi$ and $\Phi$ are the ``hidden antibaryons'' that comprise
the dark matter.
Furthermore, there exists a physical CP-violating phase
$\arg(\lambda_1^* \lambda_2 \zeta_1 \zeta_2^*)$ that cannot be
removed through phase redefinitions of the fields.

  We also introduce a hidden $U(1)^\prime$ gauge symmetry under
which $\Psi$ and $\Phi$ have opposite charges $\pm e^\prime$,
while $X_a$ is neutral.  We assume this symmetry is spontaneously
broken at the GeV scale,
and has a kinetic mixing with SM hypercharge $U(1)_Y$ via the coupling
$-\frac{\kappa}{2}B_{\mu\nu}Z'_{\mu\nu}$,
where $B_{\mu \nu}$ and $Z^\prime_{\mu\nu}$ are the $U(1)_Y$ and $U(1)'$
field strength tensors.  At energies well below the electroweak scale
the effect of this mixing is primarily to generate a vector coupling of the
massive $Z'$ gauge boson to SM particles with strength
$-c_W\kappa\,Q_{em}e$.  The $\gev$-scale $Z'$ masses
we consider here can be consistent with observations for
$10^{-6} \lesssim \kappa \lesssim 10^{-2}$~~\cite{Pospelov:2008zw,
Bjorken:2009mm}.

  In the mechanism for hylogenesis presented in Ref.~\cite{Davoudiasl:2010am},
baryogenesis begins when a non-thermal,
CP-symmetric population of $X_1$ and $\bar{X_1}$ is produced in the
early Universe.  These states decay through
$X_1 \to udd$
or
$X_1 \to \bar{\Psi}\Phi^\dagger$
(and their conjugates).
An asymmetry between the partial widths for
$X_1 \to udd $ and $\bar{X}_1\to \bar u \bar d \bar d$
arises from interference between tree and loop diagrams
and is characterized by
\bea
\epsilon = \frac{1}{2\Gamma_{X_1}}\left[\Gamma({X}_1 \to udd)
-\Gamma(\bar{X}_1\to \bar{u}\bar{d}\bar{d})\right]
\simeq
\frac{ m_{X_1}^5
\textrm{Im}[\lambda_1^* \lambda_2 \zeta_1 \zeta_2^*] }
{256 \pi^3 \, |\zeta_1|^2 \, M^4 m_{X_2}} \;,
\eea
where we have assumed that the total decay rate $\Gamma_{X_1}$ is dominated
by $X_1\to \bar{\Psi}\Phi^\dagger$ over the three-quark mode, and that
$m_{X_2} \gg m_{X_1}$.
For $\epsilon \ne 0$, $X_1$ decays generate a baryon asymmetry
in the visible sector, and by CPT an equal and opposite baryon
asymmetry in the hidden sector.  These asymmetries can be ``frozen in''
by the weakness of the coupling between both sectors provided
the temperature at which the $X_1$ are produced is not too high.
For the asymmetry to be large enough to explain the observed value,
$m_{X_{1,2}}$ cannot be too much smaller than $M$.

  Once produced, $\Psi$ and $\Phi$ will thermalize by scattering
with $Z'$ vectors present in the plasma.  These interactions will also
deplete the symmetric densities of $\Psi$ and $\Phi$ very efficiently
through annihilation to pairs of $Z'$ vectors provided
$m_{Z'} < m_{\Psi,\Phi}$~\cite{Graesser:2011wi}.
Only the asymmetries will remain.
This is analogous to the annihilation of baryons with antibaryons.
The cross section for $\Psi\bar{\Psi}\to Z'Z'$ is
given by~\cite{Pospelov:2007mp}
\bea
\langle \sigma v\rangle = \frac{{e^\prime}^4}{16\pi}\frac{1}{m_\Psi^2}
\sqrt{1-m_{Z'}^2/m_\Psi^2}
\simeq (1.6\times 10^{-25}\textrm{cm}^3/\textrm{s})\lrf{e^\prime}{0.05}^4\lrf{3\,\gev}{m_\Psi}^2.
\eea
Annihilation of $\Phi^\dagger\Phi$ is given by a similar expression.

  There is also a direct detection signal in our model due
to the hidden $Z'$ mediating the elastic scattering of
$\Psi$ and $\Phi$ off protons.
The effective scattering cross section per nucleon for either
$\Psi$ or $\Phi$ is spin-independent and given by
\bea
\sigma_0^{SI} = (5\times 10^{-39}\textrm{cm}^2)\lrf{2Z}{A}^2\lrf{\mu_{N}}{\gev}^2
\lrf{e'}{0.05}^2
\lrf{\kappa}{10^{-5}}^2
\lrf{0.1\gev}{m_{Z'}}^4,
\eea
where $\mu_N$ is the DM-nucleon reduced mass.
For a DM mass of $2.9\,\gev$, this is slightly below the best current
limits from CRESST~\cite{Angloher:2002in},
CDMS~\cite{Ahmed:2010wy}, and CoGeNT~\cite{Aalseth:2010vx}.

\end{appendix}

%%%%%%%%%%%%%%%%%%%%%%%%%%%%%%%%%%%%%%%%%%%%%%%%%%%%%%%%%%%%%%%%%%%%%%

%\newpage


\begin{thebibliography}{9}
%\begin{references}

%\cite{Komatsu:2010fb}
\bibitem{Komatsu:2010fb}
  E.~Komatsu {\it et al.} [ WMAP Collaboration ],
  %``Seven-Year Wilkinson Microwave Anisotropy Probe (WMAP) Observations: Cosmological Interpretation,''
  Astrophys.\ J.\ Suppl.\  {\bf 192}, 18 (2011).
  [1001.4538 [astro-ph.CO]].

%\cite{Riotto:1999yt}
\bibitem{Riotto:1999yt}
For reviews, see:
  A.~Riotto, M.~Trodden,
  %``Recent progress in baryogenesis,''
  Ann.\ Rev.\ Nucl.\ Part.\ Sci.\  {\bf 49}, 35-75 (1999).
  [hep-ph/9901362];
  M.~Dine, A.~Kusenko,
  %``The Origin of the matter - antimatter asymmetry,''
  Rev.\ Mod.\ Phys.\  {\bf 76}, 1 (2004).
  [hep-ph/0303065];
  S.~Davidson, E.~Nardi, Y.~Nir,
  %``Leptogenesis,''
  Phys.\ Rept.\  {\bf 466}, 105-177 (2008).
  [0802.2962 [hep-ph]].

%\cite{Jungman:1995df}
\bibitem{Jungman:1995df}
For reviews, see:\\
  G.~Jungman, M.~Kamionkowski, K.~Griest,
  %``Supersymmetric dark matter,''
  Phys.\ Rept.\  {\bf 267}, 195-373 (1996).
  [hep-ph/9506380];
%\cite{Bertone:2004pz}
%\bibitem{Bertone:2004pz}
  G.~Bertone, D.~Hooper, J.~Silk,
  %``Particle dark matter: Evidence, candidates and constraints,''
  Phys.\ Rept.\  {\bf 405}, 279-390 (2005).
  [hep-ph/0404175].


  %\cite{Nussinov:1985xr} !ADM papers
\bibitem{Nussinov:1985xr}
  S.~Nussinov,
  %``Technocosmology: Could A Technibaryon Excess Provide A 'natural' Missing Mass Candidate?,''
  Phys.\ Lett.\  {\bf B165}, 55 (1985);
%\cite{Barr:1990ca}
%\bibitem{Barr:1990ca}
  S.~M.~Barr, R.~S.~Chivukula, E.~Farhi,
  %``Electroweak Fermion Number Violation And The Production Of Stable Particles In The Early Universe,''
  Phys.\ Lett.\  {\bf B241}, 387-391 (1990).
%\cite{Barr:1991qn}
%\bibitem{Barr:1991qn}
  S.~M.~Barr,
  %``Baryogenesis, sphalerons and the cogeneration of dark matter,''
  Phys.\ Rev.\  {\bf D44}, 3062-3066 (1991);
%\cite{Kaplan:1991ah}
%\bibitem{Kaplan:1991ah}
  D.~B.~Kaplan,
  %``A Single explanation for both the baryon and dark matter densities,''
  Phys.\ Rev.\ Lett.\  {\bf 68}, 741-743 (1992);
%\cite{Kaplan:2009ag}
%\bibitem{Kaplan:2009ag}
  D.~E.~Kaplan, M.~A.~Luty, K.~M.~Zurek,
  %``Asymmetric Dark Matter,''
  Phys.\ Rev.\  {\bf D79}, 115016 (2009).
  [0901.4117 [hep-ph]];
%\cite{Kribs:2009fy}
%\bibitem{Kribs:2009fy}
  G.~D.~Kribs, T.~S.~Roy, J.~Terning, K.~M.~Zurek,
  %``Quirky Composite Dark Matter,''
  Phys.\ Rev.\  {\bf D81}, 095001 (2010).
  [0909.2034 [hep-ph]];
%\cite{Cohen:2010kn}
%\bibitem{Cohen:2010kn}
  T.~Cohen, D.~J.~Phalen, A.~Pierce, K.~M.~Zurek,
  %``Asymmetric Dark Matter from a GeV Hidden Sector,''
  Phys.\ Rev.\  {\bf D82}, 056001 (2010),
  [1005.1655 [hep-ph]].

%\cite{Hooper:2004dc}
\bibitem{Hooper:2004dc}
  D.~Hooper, J.~March-Russell, S.~M.~West,
  %``Asymmetric sneutrino dark matter and the Omega(b) / Omega(DM) puzzle,''
  Phys.\ Lett.\  {\bf B605}, 228-236 (2005).
  [hep-ph/0410114].

%\cite{Kitano:2004sv}
\bibitem{Kitano:2004sv}
  R.~Kitano, I.~Low,
  %``Dark matter from baryon asymmetry,''
  Phys.\ Rev.\  {\bf D71}, 023510 (2005).
  [hep-ph/0411133];
%\cite{Kitano:2005ge}
%\bibitem{Kitano:2005ge}
  R.~Kitano, I.~Low,
  %``Grand unification, dark matter, baryon asymmetry, and the small scale structure of the universe,''
  [hep-ph/0503112].

%\cite{Agashe:2004bm}
\bibitem{Agashe:2004bm}
  K.~Agashe, G.~Servant,
  %``Baryon number in warped GUTs: Model building and (dark matter related) phenomenology,''
  JCAP {\bf 0502}, 002 (2005).
  [hep-ph/0411254].

%\cite{Farrar:2005zd}
\bibitem{Farrar:2005zd}
  G.~R.~Farrar, G.~Zaharijas,
  %``Dark matter and the baryon asymmetry,''
  Phys.\ Rev.\ Lett.\  {\bf 96}, 041302 (2006).
  [hep-ph/0510079].

%\cite{Shelton:2010ta}
\bibitem{Shelton:2010ta}
  J.~Shelton, K.~M.~Zurek,
  %``Darkogenesis: A baryon asymmetry from the dark matter sector,''
  Phys.\ Rev.\  {\bf D82}, 123512 (2010).
  [1008.1997 [hep-ph]];
%\cite{Haba:2010bm}
%\bibitem{Haba:2010bm}
  N.~Haba, S.~Matsumoto,
  %``Baryogenesis from Dark Sector,''
  [1008.2487 [hep-ph]];
%\cite{Buckley:2010ui}
%\bibitem{Buckley:2010ui}
  M.~R.~Buckley, L.~Randall,
  %``Xogenesis,''
  [1009.0270 [hep-ph]];
%\cite{Chun:2010hz}
%\bibitem{Chun:2010hz}
  E.~J.~Chun,
  %``Leptogenesis origin of Dirac gaugino dark matter,''
  Phys.\ Rev.\  {\bf D83}, 053004 (2011).
  [1009.0983 [hep-ph]];
%\cite{McDonald:2010rn}
%\bibitem{McDonald:2010rn}
  J.~McDonald,
  %``Baryomorphosis: Relating the Baryon Asymmetry to the 'WIMP Miracle',''
  [1009.3227 [hep-ph]];
%\cite{Blennow:2010qp}
%\bibitem{Blennow:2010qp}
  M.~Blennow, B.~Dasgupta, E.~Fernandez-Martinez, N.~Rius,
  %``Aidnogenesis via Leptogenesis and Dark Sphalerons,''
  JHEP {\bf 1103}, 014 (2011),
  [1009.3159 [hep-ph]];
%\cite{Hall:2010jx}
%\bibitem{Hall:2010jx}
  L.~J.~Hall, J.~March-Russell, S.~M.~West,
  %``A Unified Theory of Matter Genesis: Asymmetric Freeze-In,''
  [1010.0245 [hep-ph]];
%\cite{Dutta:2010va}
%\bibitem{Dutta:2010va}
  B.~Dutta, J.~Kumar,
  %``Asymmetric Dark Matter from Hidden Sector Baryogenesis,''
  Phys.\ Lett.\  {\bf B699}, 364-367 (2011).
  [1012.1341 [hep-ph]];
%\cite{Falkowski:2011xh}
%\bibitem{Falkowski:2011xh}
  A.~Falkowski, J.~T.~Ruderman, T.~Volansky,
  %``Asymmetric Dark Matter from Leptogenesis,''
  JHEP {\bf 1105}, 106 (2011),
  [1101.4936 [hep-ph]];
%\cite{Chun:2011cc}
%\bibitem{Chun:2011cc}
  E.~J.~Chun,
  %``Minimal Dark Matter and Leptogenesis,''
  JHEP {\bf 1103}, 098 (2011),
  [1102.3455 [hep-ph]];
%\cite{Kang:2011wb}
%\bibitem{Kang:2011wb}
  Z.~Kang, J.~Li, T.~Li, T.~Liu, J.~Yang,
  %``Asymmetric Sneutrino Dark Matter in the NMSSM with Minimal Inverse Seesaw,''
  [1102.5644 [hep-ph]];
%\cite{Heckman:2011sw}
%\bibitem{Heckman:2011sw}
  J.~J.~Heckman, S.~-J.~Rey,
  %``Baryon and Dark Matter Genesis from Strongly Coupled Strings,''
  [1102.5346 [hep-th]];
%\cite{Kaplan:2011yj}
%\bibitem{Kaplan:2011yj}
  D.~E.~Kaplan, G.~Z.~Krnjaic, K.~R.~Rehermann, C.~M.~Wells,
  %``Dark Atoms: Asymmetry and Direct Detection,''
  [1105.2073 [hep-ph]].
%\cite{Frandsen:2011kt}
%\bibitem{Frandsen:2011kt}
  M.~T.~Frandsen, S.~Sarkar, K.~Schmidt-Hoberg,
  %``Light asymmetric dark matter from new strong dynamics,''
  [1103.4350 [hep-ph]];
%\cite{Hook:2011tk}
%\bibitem{Hook:2011tk}
  A.~Hook,
  %``Unitarity constraints on asymmetric freeze-in,''
  [1105.3728 [hep-ph]];
%\cite{Bell:2011tn}
%\bibitem{Bell:2011tn}
  N.~F.~Bell, K.~Petraki, I.~M.~Shoemaker, R.~R.~Volkas,
  %``Pangenesis in a Baryon-Symmetric Universe: Dark and Visible Matter via the Affleck-Dine Mechanism,''
  [1105.3730 [hep-ph]];
%\cite{Cheung:2011if}
%\bibitem{Cheung:2011if}
  C.~Cheung, K.~M.~Zurek,
  %``Affleck-Dine Cogenesis,''
  [1105.4612 [hep-ph]].

%\cite{Hut:1979xw}
\bibitem{Hut:1979xw}
  P.~Hut, K.~A.~Olive,
  %``A Cosmological Upper Limit On The Mass Of Heavy Neutrinos,''
  Phys.\ Lett.\  {\bf B87}, 144-146 (1979).

%\cite{Dodelson:1989cq}
\bibitem{Dodelson:1989cq}
  S.~Dodelson, L.~M.~Widrow,
  %``Baryogenesis In A Baryon Symmetric Universe,''
  Phys.\ Rev.\  {\bf D42}, 326-342 (1990).

%\cite{Kuzmin:1996he}
\bibitem{Kuzmin:1996he}
  V.~A.~Kuzmin,
  %``A Simultaneous solution to baryogenesis and dark matter problems,''
  Phys.\ Part.\ Nucl.\  {\bf 29}, 257-265 (1998).
  [hep-ph/9701269].

%\cite{Gu:2007cw}
\bibitem{Gu:2007cw}
  P.~-H.~Gu,
  %``Origin of matter in the universe,''
  Phys.\ Lett.\  {\bf B657}, 103-106 (2007).
  [0706.1946 [hep-ph]];
%\cite{Gu:2009yy}
%\bibitem{Gu:2009yy}
  P.~-H.~Gu, U.~Sarkar, X.~Zhang,
  %``Visible and Dark Matter Genesis and Cosmic Positron/Electron Excesses,''
  Phys.\ Rev.\  {\bf D80}, 076003 (2009).
  [0906.3103 [hep-ph]];
%\cite{Gu:2010ft}
%\bibitem{Gu:2010ft}
  P.~-H.~Gu, M.~Lindner, U.~Sarkar, X.~Zhang,
  %``WIMP Dark Matter and Baryogenesis,''
  [1009.2690 [hep-ph]].

%\cite{An:2009vq}
\bibitem{An:2009vq}
  H.~An, S.~-L.~Chen, R.~N.~Mohapatra, Y.~Zhang,
  %``Leptogenesis as a Common Origin for Matter and Dark Matter,''
  JHEP {\bf 1003}, 124 (2010).
  [0911.4463 [hep-ph]];
%\cite{An:2010kc}
%\bibitem{An:2010kc}
  H.~An, S.~-L.~Chen, R.~N.~Mohapatra, S.~Nussinov, Y.~Zhang,
  %``Energy Dependence of Direct Detection Cross Section for Asymmetric Mirror Dark Matter,''
  Phys.\ Rev.\  {\bf D82}, 023533 (2010).
  [1004.3296 [hep-ph]].

%\cite{Davoudiasl:2010am}
\bibitem{Davoudiasl:2010am}
  H.~Davoudiasl, D.~E.~Morrissey, K.~Sigurdson, S.~Tulin,
  %``Hylogenesis: A Unified Origin for Baryonic Visible Matter and Antibaryonic Dark Matter,''
  Phys.\ Rev.\ Lett.\  {\bf 105}, 211304 (2010).
  [1008.2399 [hep-ph]].

%\cite{Chung:2008gv}
\bibitem{Chung:2008gv}
  D.~J.~H.~Chung, B.~Garbrecht, S.~Tulin,
  %``The Effect of the Sparticle Mass Spectrum on the Conversion of B-L to B,''
  JCAP {\bf 0903}, 008 (2009).
  [0807.2283 [hep-ph]].

%\cite{Dimopoulos:1987rk}
\bibitem{Dimopoulos:1987rk}
  S.~Dimopoulos, L.~J.~Hall,
  %``Baryogenesis At The Mev Era,''
  Phys.\ Lett.\  {\bf B196}, 135 (1987);
%\cite{Cline:1990bw}
%\bibitem{Cline:1990bw}
  J.~M.~Cline, S.~Raby,
  %``Gravitino induced baryogenesis: A Problem made a virtue,''
  Phys.\ Rev.\  {\bf D43}, 1781-1787 (1991);
%\cite{Thomas:1995ze}
%\bibitem{Thomas:1995ze}
  S.~D.~Thomas,
  %``Baryons and dark matter from the late decay of a supersymmetric condensate,''
  Phys.\ Lett.\  {\bf B356}, 256-263 (1995).
  [hep-ph/9506274];
%\cite{Kitano:2008tk}
%\bibitem{Kitano:2008tk}
  R.~Kitano, H.~Murayama, M.~Ratz,
  %``Unified origin of baryons and dark matter,''
  Phys.\ Lett.\  {\bf B669}, 145-149 (2008).
  [0807.4313 [hep-ph]];
%\cite{Allahverdi:2010rh}
%\bibitem{Allahverdi:2010rh}
  R.~Allahverdi, B.~Dutta, K.~Sinha,
  %``Cladogenesis: Baryon-Dark Matter Coincidence from Branchings in Moduli Decay,''
  Phys.\ Rev.\  {\bf D83}, 083502 (2011).
  [1011.1286 [hep-ph]].


%\cite{Kobayashi:2005pe}
\bibitem{Kobayashi:2005pe}
  K.~Kobayashi {\it et al.} [ Super-Kamiokande Collaboration ],
  %``Search for nucleon decay via modes favored by supersymmetric grand unification models in Super-Kamiokande-I,''
  Phys.\ Rev.\  {\bf D72}, 052007 (2005).
  [hep-ex/0502026].

%\cite{Nath:2006ut}
\bibitem{Nath:2006ut}
  For a review of various aspects of nucleon stability within
  unified models, see, for example, P.~Nath, P.~Fileviez Perez,
  %``Proton stability in grand unified theories, in strings and in branes,''
  Phys.\ Rept.\  {\bf 441}, 191-317 (2007).
  [hep-ph/0601023].


%\cite{Wall:2000pq}
\bibitem{Wall:2000pq}
  D.~Wall {\it et al.} [ Soudan 2 Collaboration ],
  %``Search for nucleon decay with final states lepton+ eta**0, anti-neutrino eta**0, and anti-neutrino pi+,0 using Soudan-2,''
  Phys.\ Rev.\  {\bf D62}, 092003 (2000).
  [hep-ex/0001015].

%\cite{McGrew:1999nd}
\bibitem{McGrew:1999nd}
  C.~McGrew, R.~Becker-Szendy, C.~B.~Bratton, J.~L.~Breault, D.~R.~Cady, D.~Casper, S.~T.~Dye, W.~Gajewski {\it et al.},
  %``Search for nucleon decay using the IMB-3 detector,''
  Phys.\ Rev.\  {\bf D59}, 052004 (1999).

%\cite{McGrew:1994nd}
\bibitem{McGrew:1994nd}
  C.~D.~McGrew,
  ``A Search for baryon nonconservation using the IMB-3 Detector,'' Ph.~D.~thesis,
  University of California at Irvine,. 1994.

%\cite{Claudson:1981gh}
\bibitem{Claudson:1981gh}
  M.~Claudson, M.~B.~Wise and L.~J.~Hall,
  %``Chiral Lagrangian For Deep Mine Physics,''
  Nucl.\ Phys.\  B {\bf 195}, 297 (1982).
  %%CITATION = NUPHA,B195,297;%

%\cite{Aoki:1999tw}
\bibitem{Aoki:1999tw}
  S.~Aoki {\it et al.} [ JLQCD Collaboration ],
  %``Nucleon decay matrix elements from lattice QCD,''
  Phys.\ Rev.\  {\bf D62}, 014506 (2000).
  [hep-lat/9911026].

%\cite{Suzuki:2001rb}
\bibitem{Suzuki:2001rb}
  Y.~Suzuki {\it et al.} [ TITAND Working Group Collaboration ],
  %``Multimegaton water Cherenkov detector for a proton decay search: TITAND (former name TITANIC),''
[hep-ex/0110005].

%\cite{Diwan:2003uw}
\bibitem{Diwan:2003uw}
  M.~V.~Diwan, R.~L.~Hahn, W.~Marciano, B.~Viren, R.~Svoboda, W.~Frati, K.~Lande, A.~K.~Mann {\it et al.},
  %``Megaton modular multipurpose neutrino detector for a program of physics in the homestake DUSEL,''
[hep-ex/0306053].


%\cite{Bueno:2007um}
\bibitem{Bueno:2007um}
  A.~Bueno, Z.~Dai, Y.~Ge, M.~Laffranchi, A.~J.~Melgarejo, A.~Meregaglia, S.~Navas, A.~Rubbia,
  %``Nucleon decay searches with large liquid argon TPC detectors at shallow depths: Atmospheric neutrinos and cosmogenic backgrounds,''
  JHEP {\bf 0704}, 041 (2007).
  [hep-ph/0701101].



%\cite{Birkedal:2004xn}
\bibitem{Birkedal:2004xn}
  A.~Birkedal, K.~Matchev, M.~Perelstein,
  %``Dark matter at colliders: A Model independent approach,''
  Phys.\ Rev.\  {\bf D70}, 077701 (2004).
  [hep-ph/0403004].

%\cite{Beltran:2010ww}
\bibitem{Beltran:2010ww}
  M.~Beltran, D.~Hooper, E.~W.~Kolb, Z.~A.~C.~Krusberg, T.~M.~P.~Tait,
  %``Maverick dark matter at colliders,''
  JHEP {\bf 1009}, 037 (2010).
  [1002.4137 [hep-ph]].

%\cite{Goodman:2010yf}
\bibitem{Goodman:2010yf}
  J.~Goodman, M.~Ibe, A.~Rajaraman, W.~Shepherd, T.~M.~P.~Tait, H.~-B.~Yu,
  %``Constraints on Light Majorana dark Matter from Colliders,''
  Phys.\ Lett.\  {\bf B695}, 185-188 (2011).
  [1005.1286 [hep-ph]];
%\cite{Goodman:2010ku}
%\bibitem{Goodman:2010ku}
  J.~Goodman, M.~Ibe, A.~Rajaraman, W.~Shepherd, T.~M.~P.~Tait, H.~-B.~Yu,
  %``Constraints on Dark Matter from Colliders,''
  Phys.\ Rev.\  {\bf D82}, 116010 (2010).
  [1008.1783 [hep-ph]].

%\cite{Bai:2010hh}
\bibitem{Bai:2010hh}
  Y.~Bai, P.~J.~Fox, R.~Harnik,
  %``The Tevatron at the Frontier of Dark Matter Direct Detection,''
  JHEP {\bf 1012}, 048 (2010).
  [1005.3797 [hep-ph]].

%\cite{Fox:2011fx}
\bibitem{Fox:2011fx}
  P.~J.~Fox, R.~Harnik, J.~Kopp, Y.~Tsai,
  %``LEP Shines Light on Dark Matter,''
  [1103.0240 [hep-ph]].

%\cite{Buckley:2011kk}
\bibitem{Buckley:2011kk}
  M.~R.~Buckley,
  %``Asymmetric Dark Matter and Effective Operators,''
  [1104.1429 [hep-ph]].

%\cite{Pumplin:2002vw}
\bibitem{Pumplin:2002vw}
  J.~Pumplin, D.~R.~Stump, J.~Huston, H.~L.~Lai, P.~M.~Nadolsky, W.~K.~Tung,
  %``New generation of parton distributions with uncertainties from global QCD analysis,''
  JHEP {\bf 0207}, 012 (2002).
  [hep-ph/0201195].

%\cite{Aaltonen:2008hh}
\bibitem{Aaltonen:2008hh}
  T.~Aaltonen {\it et al.} [ CDF Collaboration ],
  %``Search for large extra dimensions in final states containing one photon or jet and large missing transverse energy produced in $p \bar{p}$ collisions at $\sqrt{s}$ = 1.96-TeV,''
  Phys.\ Rev.\ Lett.\  {\bf 101}, 181602 (2008).
  [0807.3132 [hep-ex]].

%cite{cdfmono}
\bibitem{cdfmono}
http://www-cdf.fnal.gov/physics/exotic/r2a/20070322.monojet/public/ykk.html

%\cite{Vacavant:2001sd}
\bibitem{Vacavant:2001sd}
  L.~Vacavant, I.~Hinchliffe,
  %``Signals of models with large extra dimensions in ATLAS,''
  J.\ Phys.\ G {\bf G27}, 1839-1850 (2001).

%\cite{Press:1985ug}
\bibitem{Press:1985ug}
  W.~H.~Press, D.~N.~Spergel,
  %``Capture by the sun of a galactic population of weakly interacting massive particles,''
  Astrophys.\ J.\  {\bf 296}, 679-684 (1985).

%\cite{Griest:1986yu}
\bibitem{Griest:1986yu}
  K.~Griest, D.~Seckel,
  %``Cosmic Asymmetry, Neutrinos and the Sun,''
  Nucl.\ Phys.\  {\bf B283}, 681 (1987).

%\cite{Gould:1987ir}
\bibitem{Gould:1987ir}
  A.~Gould,
  %``Resonant Enhancements in WIMP Capture by the Earth,''
  Astrophys.\ J.\  {\bf 321}, 571 (1987).

%\cite{Angloher:2002in}
\bibitem{Angloher:2002in}
  G.~Angloher {\it et al.},
  %``Limits on WIMP dark matter using sapphire cryogenic detectors,''
  Astropart.\ Phys.\  {\bf 18}, 43 (2002).
  %%CITATION = APHYE,18,43;%%

%\cite{Ahmed:2010wy}
\bibitem{Ahmed:2010wy}
  Z.~Ahmed {\it et al.} [ CDMS-II Collaboration ],
  %``Results from a Low-Energy Analysis of the CDMS II Germanium Data,''
  Phys.\ Rev.\ Lett.\  {\bf 106}, 131302 (2011).
  [1011.2482 [astro-ph.CO]].

%\cite{Aalseth:2010vx}
\bibitem{Aalseth:2010vx}
  C.~E.~Aalseth {\it et al.} [ CoGeNT Collaboration ],
  %``Results from a Search for Light-Mass Dark Matter with a P-type Point Contact Germanium Detector,''
  Phys.\ Rev.\ Lett.\  {\bf 106}, 131301 (2011).
  [1002.4703 [astro-ph.CO]].



%\cite{Bertone:2007ae}
\bibitem{Bertone:2007ae}
  G.~Bertone, M.~Fairbairn,
  %``Compact Stars as Dark Matter Probes,''
  Phys.\ Rev.\  {\bf D77}, 043515 (2008).
  [0709.1485 [astro-ph]].

%\cite{Gould:1987ju}
\bibitem{Gould:1987ju}
  A.~Gould,
  %``Wimp Distribution In And Evaporation From The Sun,''
  Astrophys.\ J.\  {\bf 321}, 560 (1987).

%\cite{Hebeler:2010jx}
\bibitem{Hebeler:2010jx}
  K.~Hebeler, J.~M.~Lattimer, C.~J.~Pethick, A.~Schwenk,
  %``Constraints on neutron star radii based on chiral effective field theory interactions,''
  Phys.\ Rev.\ Lett.\  {\bf 105}, 161102 (2010).
  [1007.1746 [nucl-th]].

%\cite{Goldman:1989nd}
\bibitem{Goldman:1989nd}
  I.~Goldman, S.~Nussinov,
  %``Weakly Interacting Massive Particles And Neutron Stars,''
  Phys.\ Rev.\  {\bf D40}, 3221-3230 (1989).

%\cite{Kouvaris:2007ay}
\bibitem{Kouvaris:2007ay}
  C.~Kouvaris,
  %``WIMP Annihilation and Cooling of Neutron Stars,''
  Phys.\ Rev.\  {\bf D77}, 023006 (2008).
  [0708.2362 [astro-ph]];
%\cite{Kouvaris:2010vv}
%\bibitem{Kouvaris:2010vv}
  C.~Kouvaris, P.~Tinyakov,
  %``Can Neutron stars constrain Dark Matter?,''
  Phys.\ Rev.\  {\bf D82}, 063531 (2010).
  [1004.0586 [astro-ph.GA]].

%\cite{deLavallaz:2010wp}
\bibitem{deLavallaz:2010wp}
  A.~de Lavallaz, M.~Fairbairn,
  %``Neutron Stars as Dark Matter Probes,''
  Phys.\ Rev.\  {\bf D81}, 123521 (2010).
  [1004.0629 [astro-ph.GA]].

%\cite{Kouvaris:2010jy}
\bibitem{Kouvaris:2010jy}
  C.~Kouvaris, P.~Tinyakov,
  %``Constraining Asymmetric Dark Matter through observations of compact stars,''
  Phys.\ Rev.\  {\bf D83}, 083512 (2011).
  [1012.2039 [astro-ph.HE]];
%\cite{Kouvaris:2011fi}
%\bibitem{Kouvaris:2011fi}
  C.~Kouvaris, P.~Tinyakov,
  %``Excluding Light Asymmetric Bosonic Dark Matter,''
  [1104.0382 [astro-ph.CO]].

%\cite{McDermott:2011jp}
\bibitem{McDermott:2011jp}
  S.~D.~McDermott, H.~-B.~Yu, K.~M.~Zurek,
  %``Constraints on Scalar Asymmetric Dark Matter from Black Hole Formation in Neutron Stars,''
  [1103.5472 [hep-ph]].

%\cite{Gould:1989gw}
\bibitem{Gould:1989gw}
  A.~Gould, B.~T.~Draine, R.~W.~Romani, S.~Nussinov,
  %``Neutron Stars: Graveyard Of Charged Dark Matter,''
  Phys.\ Lett.\  {\bf B238}, 337 (1990).

%\cite{Kepler:2006ns}
\bibitem{Kepler:2006ns}
  S.~O.~Kepler, S.~J.~Kleinman, A.~Nitta, D.~Koester, B.~G.~Castanheira, O.~Giovannini, A.~F.~M.~Costa, L.~Althaus,
  %``White Dwarf Mass Distribution in the SDSS,''
  Mon.\ Not.\ Roy.\ Astron.\ Soc.\  {\bf 375}, 1315-1324 (2007).
  [astro-ph/0612277].

%\cite{Bottino:2002pd}
\bibitem{Bottino:2002pd}
  A.~Bottino, G.~Fiorentini, N.~Fornengo, B.~Ricci, S.~Scopel, F.~L.~Villante,
  %``Does solar physics provide constraints to weakly interacting massive particles?,''
  Phys.\ Rev.\  {\bf D66}, 053005 (2002).
  [hep-ph/0206211].

%\cite{McCullough:2010ai}
\bibitem{McCullough:2010ai}
  M.~McCullough, M.~Fairbairn,
  %``Capture of Inelastic Dark Matter in White Dwarves,''
  Phys.\ Rev.\  {\bf D81}, 083520 (2010).
  [1001.2737 [hep-ph]].

%\cite{Hooper:2010es}
\bibitem{Hooper:2010es}
  D.~Hooper, D.~Spolyar, A.~Vallinotto, N.~Y.~Gnedin,
  %``Inelastic Dark Matter As An Efficient Fuel For Compact Stars,''
  Phys.\ Rev.\  {\bf D81}, 103531 (2010).
  [1002.0005 [hep-ph]].

%\cite{Hooper:2008cf}
\bibitem{Hooper:2008cf}
  D.~Hooper, F.~Petriello, K.~M.~Zurek, M.~Kamionkowski,
  %``The New DAMA Dark-Matter Window and Energetic-Neutrino Searches,''
  Phys.\ Rev.\  {\bf D79}, 015010 (2009).
  [0808.2464 [hep-ph]].

%\cite{Frandsen:2010yj}
\bibitem{Frandsen:2010yj}
  M.~T.~Frandsen, S.~Sarkar,
  %``Asymmetric dark matter and the Sun,''
  Phys.\ Rev.\ Lett.\  {\bf 105}, 011301 (2010).
  [1003.4505 [hep-ph]];
%\cite{Cumberbatch:2010hh}
%\bibitem{Cumberbatch:2010hh}
  D.~T.~Cumberbatch, J.~.A.~Guzik, J.~Silk, L.~S.~Watson, S.~M.~West,
  %``Light WIMPs in the Sun: Constraints from Helioseismology,''
  Phys.\ Rev.\  {\bf D82}, 103503 (2010).
  [1005.5102 [astro-ph.SR]];
%\cite{Taoso:2010tg}
%\bibitem{Taoso:2010tg}
  M.~Taoso, F.~Iocco, G.~Meynet, G.~Bertone, P.~Eggenberger,
  %``Effect of low mass dark matter particles on the Sun,''
  Phys.\ Rev.\  {\bf D82}, 083509 (2010).
  [1005.5711 [astro-ph.CO]];
%\cite{Casanellas:2010he}
%\bibitem{Casanellas:2010he}
  J.~Casanellas, I.~Lopes,
  %``Towards the use of asteroseismology to investigate the nature of dark matter,''
  Mon.\ Not.\ Roy.\ Astron.\ Soc.\  {\bf 410}, 535-540 (2011).
  [1008.0646 [astro-ph.CO]].
%\cite{Lopes:2010fx}
%\bibitem{Lopes:2010fx}
  I.~Lopes, J.~Silk,
  %``Probing the Existence of a Dark Matter Isothermal Core Using Gravity Modes,''
  Astrophys.\ J.\  {\bf 722}, L95 (2010).
  [1009.5122 [astro-ph.SR]].

%\cite{Abbasi:2009uz}
\bibitem{Abbasi:2009uz}
  R.~Abbasi {\it et al.} [ ICECUBE Collaboration ],
  %``Limits on a muon flux from neutralino annihilations in the Sun with the IceCube 22-string detector,''
  Phys.\ Rev.\ Lett.\  {\bf 102}, 201302 (2009).
  [0902.2460 [astro-ph.CO]].

%\cite{Fairbairn:2007bn}
\bibitem{Fairbairn:2007bn}
  M.~Fairbairn, P.~Scott, J.~Edsjo,
  %``The Zero Age Main Sequence of WIMP burners,''
  Phys.\ Rev.\  {\bf D77}, 047301 (2008).
  [0710.3396 [astro-ph]];
%\cite{Scott:2008ns}
%\bibitem{Scott:2008ns}
  P.~Scott, M.~Fairbairn, J.~Edsjo,
  %``Dark stars at the Galactic centre - the main sequence,''
  Mon.\ Not.\ Roy.\ Astron.\ Soc.\  {\bf 394}, 82 (2008).
  [0809.1871 [astro-ph]].



%\cite{Pospelov:2008zw}
\bibitem{Pospelov:2008zw}
  M.~Pospelov,
  %``Secluded U(1) below the weak scale,''
  Phys.\ Rev.\  D {\bf 80}, 095002 (2009).
  %%CITATION = PHRVA,D80,095002;%%

%\cite{Bjorken:2009mm}
\bibitem{Bjorken:2009mm}
  J.~D.~Bjorken, R.~Essig, P.~Schuster and N.~Toro,
  %``New Fixed-Target Experiments to Search for Dark Gauge Forces,''
  Phys.\ Rev.\  D {\bf 80}, 075018 (2009)
  [0906.0580 [hep-ph]].
  %%CITATION = PHRVA,D80,075018;%%

%\cite{Graesser:2011wi}
\bibitem{Graesser:2011wi}
  M.~L.~Graesser, I.~M.~Shoemaker, L.~Vecchi,
  %``Asymmetric WIMP dark matter,''
  [1103.2771 [hep-ph]];
%\cite{Iminniyaz:2011yp}
%\bibitem{Iminniyaz:2011yp}
  H.~Iminniyaz, M.~Drees, X.~Chen,
  %``Relic Abundance of Asymmetric Dark Matter,''
  [1104.5548 [hep-ph]].

%\cite{Pospelov:2007mp}
\bibitem{Pospelov:2007mp}
  M.~Pospelov, A.~Ritz and M.~B.~Voloshin,
  %``Secluded WIMP Dark Matter,''
  Phys.\ Lett.\  B {\bf 662}, 53 (2008).
  %%CITATION = PHLTA,B662,53;%%



\end{thebibliography}
\end{document}